\documentclass[aps,prl,preprint,superscriptaddress]{revtex4-2}

\usepackage{amsmath}
\usepackage{amssymb}
\usepackage{graphicx}
\usepackage{color}
\usepackage{makecell}
\usepackage{booktabs}
\usepackage{floatrow}
\usepackage{notes2bib}

\begin{document}


\title{Decoding the Complexity of Ferroelectric Orthorhombic $\mathrm{HfO_2}$: A Unified Mode Expansion Approach}

\author{Chenxi Yu}
\author{Jiajia Zhang}
\author{Xujin Song}

\author{Fei Liu}
\email[]{feiliu@pku.edu.cn}

\author{Jinfeng Kang}
\email[]{kangjf@pku.edu.cn}
\affiliation{School of Integrated Circuits, Peking University, Beijing, China}


\begin{abstract}
    The ferroelectricity in $\mathrm{HfO}_2$ thin films is widely attributed to the formation of a polar orthorhombic phase named OIII phase. 
	However, the complexity of OIII phase originated from its low symmetry becomes an obstacle for studying ferroelectric properties of $\mathrm{HfO}_2$. 
	Here, we developed a unified framework based on phonon mode expansion for studying ferroelectric $\mathrm{HfO}_2$. 
	In this framework, phase structures, domain walls and switching paths of orthogonal crystal system can be studied from the same basis of mode analysis. The OIII phase and other orthogonal phases can be represented by the high-symmetry cubic phase with the excitation of cubic phonon modes, into which the complexity of orthogonal phases is faithfully coded. 
	To present the capability of this mode expansion approach, we clarified the origin of orthorhombic stability from the energy functional of modes; enumerated inequivalent domain walls and calculated their stable criteria; and summarized all possible switching mechanisms. 
	This unified framework can be used to simplify the study of domain wall structures and transition paths. Furthermore, it can provide a new perspective for ferroelectricity in $\mathrm{HfO}_2$ from phonon mode analysis. 
\end{abstract}


\maketitle

\section{Introduction}

The $\mathrm{HfO}_2$-based ferroelectrics are emerging ferroelectric materials promising for non-volatile memory and compute-in memory devices in artificial intelligence (AI) applications. 
Since its discovery in 2011 \citep{RN71}, it has attracted extensive research interests for its CMOS-compatibility \citep{RN86,RN149,RN76} (complementary metal oxide semiconductor), strong ferroelectricity in nano-scale \citep{RN149,RN77,RN74,RN87} which make it promising for CMOS technology. 
The origin of ferroelectricity in $\mathrm{HfO}_2$ thin films is widely acknowledged to be the formation of a ferroelectric orthorhombic phase, OIII phase \citep{RN60}. 
The ferroelectric switching path of OIII phase is the tetragonal path with the tetragonal phase (T phase) as the intermediate state \citep{RN60}. 
However, the switching path of OIII phase is not unique. Ma \textit{et al.} found that the switching barrier depends on the $X_2^\prime$ phonon mode \citep{RN160}, and the moving directions of oxygen atoms \citep{RN251}. 
This phenomenon also exists in the domain walls (DWs) of OIII phase. 
The DWs of ferroelectric $\mathrm{HfO}_2$ are not uniquely determined by the polarization directions. These inequivalent DWs with the same polarization directions have different stabilities \citep{RN357,RN356} and switching barriers \citep{RN349,RN34,RN356}, making the classification of inequivalent DWs necessary. 
The inequivalent DWs with the same polarization directions are caused by the eight variants of OIII unitcells with the same polarization. To represent these variants of OIII unitcells, the concept of pseudo-chirality is introduced \citep{RN34,RN365}. The pseudo-chirality simplified the classification of inequivalent DWs, however, the detailed notations used are different between the works by different researchers, and many works did not take the pseudo-chirality of OIII phase into consideration when dealing with domain structures or switching paths because of its complexity, consequently missing some of the inequivalent domain walls or switching paths. 
Specificly, there are few works on switching path that used the concept of pseudo-chirality, probably due to the fact that the relation between transition path and the unitcell variants is not obvious. 
In addition, there are other phases that have variants of unitcells, such as the Pbcn phase \citep{RN360}. The number of unitcell variants depends on the nature of crystalline phases, which cannot be represented by the pseudo-chirality of OIII phase. 
To solve these issues, a key feature that can decode the complexity of both phase structures, domain walls and transition paths is required. This feature should also be physical to avoid the arbitrary choice of notations. 
The OIII unitcell variants can be transformed to each other by the symmetry group of cubic space group $Fm\overline{3}m$, therefore, the feature should also be transformed by the cubic space group. 
In addition, the feature should be able to represent superstructures including domain walls, and the deformed structures along the ferroelectric transition path. 
To find the feature with above mentioned properties, we reviewed the theory describing the ferroelectricity in perovskite ferroelectrics. In the soft-mode theory, the ferroelectric transition is caused by the softening of soft optic phonon modes. 
The effective hamiltonian theory \citep{zhong1995first} of $\mathrm{BaTiO}_3$ marks the triumph of soft-mode theory. The hamiltonian is expanded up to the fourth order by the local soft modes and strain, which is effective for it neglects higher order self-energy and interactions. Almost all the physical properties and dynamics of ferroelectrics can be simulated once we can model the hamiltonian of this material. 
Although we cannot apply the effective hamiltonian theory directly to the $\mathrm{HfO}_2$ ferroelectrics due to the complicated energy landscape, we can still apply some of the methods used in the theory to $\mathrm{HfO}_2$-based ferroelectrics. 
The phonon mode localized in each unitcell is the real space representation of phonon, which is the intrinsic property of $\mathrm{HfO}_2$ phase. 
The phase transition is accompanied by the evolution of mode amplitudes, which means that we can track the transition process by tracking the evolution of mode amplitudes. 
The localized nature of these local modes is suitable for the construction of superstructures such as DWs. 
The phonon modes transform by corresponding irreducible representations of crystalline symmetry groups, therefore, they can represent the symmetry properties of phase unitcells. 
Furthermore, the group-subgroup relation of high-symmetry non-ferroelectric phase and low-symmetry ferroelectric phase reflects the difference in the number of unitcell variants between phases: the phase with lower symmetry has more unitcell variants. 
Based on the above-mentioned facts, we developed a unified framework based on mode expansion for the studying of ferroelectric $\mathrm{HfO}_2$. 
We proposed that the phonon mode localized in each unitcell is the key feature that can represent the pseudo-chirality of OIII phase, the inequivalent DWs caused by pseudo-chirality of each domains and the different transition paths with the same initial and final states. 
We chose the cubic phase, C phase with space group $Fm\overline{3}m$ as the parent phase, or the reference phase whose symmetry group is the super group of all $\mathrm{HfO}_2$ phases that are frequently studied theoretically or observed experimentally \citep{RN60,RN369}. 
The phonon spectrum of C phase has been studied theoretically \citep{RN4,RN135}, from which we can calculate the phonon modes of C phase. 
After we obtained the phonon modes, we expanded the atom displacements in each unitcell relative to the atomic coordinates in the C phase unitcell by the phonon modes of C phase. 
A complete basis of phonon modes is required, not only because neglecting some modes may raise the symmetry, but also due to the fact that ferroelectric transition involves both soft modes and hard modes in $\mathrm{HfO}_2$ ferroelectrics \citep{RN349,RN365,RN37}. 
The strain can be represented by the strain tensor. 
The reverse process to the mode expansion is also available: the unitcells, DWs and transition paths can be generated from the corresponding C phase version by adding the phonon modes and exerting strain from strain tensor. 
By expanding into phonon modes, the symmetry properties of both phase structures, DWs and transition paths can be studied by studying the symmetry of modes. 
To demonstrate the capability of this mode expansion approach, we studied the stability and pseudo-chirality of ferroelectric orthorhombic phase (OIII) by the energy functional of mode amplitudes. 
Using the symmetry of modes, we enumerated all inequivalent domain wall structures and studied the dependence of domain wall stability on interface modes using first-principle calculations. Finally, we enumerated inequivalent ferroelectric switching paths by analysing the symmetry of modes, summarized the switching mechanisms of OIII phase and studied the role of modes on ferroelectric switching. 

\section{Phonon Mode Analysis of Cubic Phase}

\begin{figure}
    \includegraphics[scale=1]{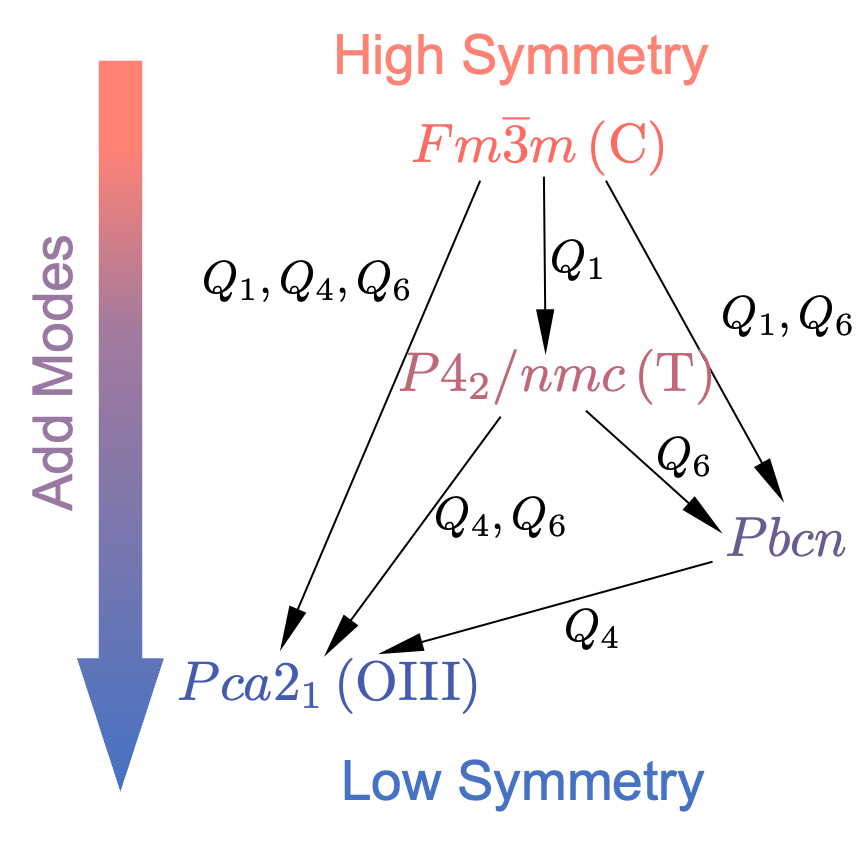}
    \caption{Group-subgroup relation of $\mathrm{HfO}_2$ phases. The low symmetry phase can be seen as the high symmetry phase superimposed by phonon modes and strained by strain tensor. 
    \label{fig:group-subgroup-relation}}
\end{figure}

\begin{figure}
    \includegraphics[scale=1]{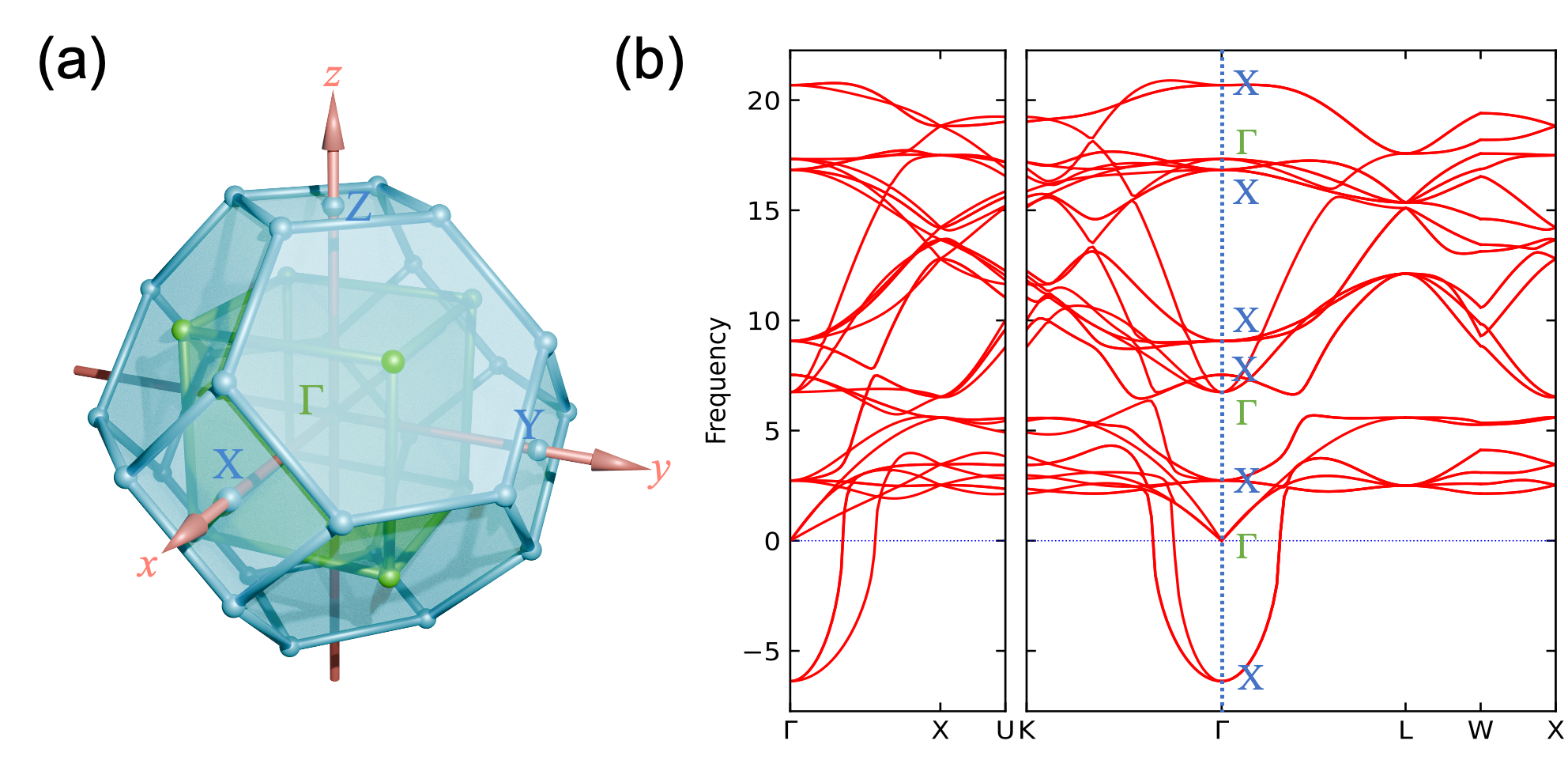}
    \caption{Phonon spectrum of $\mathrm{HfO}_2$ using conventional cell. 
    (a) The Brillouin zone of C phase primitive cell in blue and that of conventional cell in green. Due to band folding, $\Gamma$, X, Y, Z points of primitive cell are folded into the $\Gamma$ point of conventional cell. 
	(b) Phonon spectrum of C phase conventional cell. The $\Gamma$, $X$ symbols on the wavy line indicate the k points of these phonons in reciprocal space of primitive cells. 
    \label{fig:phonon-spectrum}}
\end{figure}

Analyzing the group-subgroup relations of $\mathrm{HfO}_2$ phases \citep{RN369}, we found that all the space groups of phases belonging to orthogonal crystal system are the subgroup of cubic space group $Fm\overline{3}m$ (Fig. \ref{fig:group-subgroup-relation}). 
For symmetry considerations, we chose the cubic $Fm\overline{3}m$ phase as the parent phase, or reference phase, to make mode expansion as previous works \cite{RN60}. 
Other reference phases are possible \citep{RN329}, but the low symmetry of these phases introduce complications into our analysis. 
$Fm\overline{3}m$ phase has a face-centered cubic (FCC) lattice. 
We will use the cubic conventional cell throughout the article, because of its high-symmetry and the ease to build supercell structures from unitcells. 
Moreover, the phases closely related with ferroelectric transition, including tetragonal ($P4_2/nmc$) and ferroelectric orthorhombic ($Pca2_1$) phases, can be generated from parent cubic phase by the collective motion of local phonon modes in conventional cell, or the phonons on the $\Gamma$ point in reciprocal space of the conventional cell. 
While in primitive cell, we are required to add the phonons on the X, Y and Z points which introduces complications. 
Fig. \ref{fig:phonon-spectrum}b shows the phonon spectrum of cubic $Fm\overline{3}m$ phase. 
Due to band folding, the X (100), Y (010) and Z (001) points of the primitive cell are folded into the $\Gamma$ point of the conventional cell. 
On the phonon band graph (Fig. \ref{fig:phonon-spectrum}b), we label the folded phonons at the $\Gamma$ point of conventional cell with the original k points from the primitive cell. 
At the $\Gamma$ point, the phonon spectrum comprises nine phonon branches with varying degeneracies, accounting for total of 36 vibrational degrees of freedom. 
These 36 degrees of freedom originate from the three displacement degrees of freedom of four Hf atoms and eight oxygen atoms in one conventional unitcell. 
The degeneracy of each phonon branch arises from two distinct sources, the intrinsic dimension of its irreducible representation dictated by group representation theory, and the extrinsic geometry-driven degeneracy resulting from Brillouin zone folding. 
The zone-folding degeneracy multiplicity is determined by the number of symmetrically equivalent wavevectors in the primitive Brillouin zone. 
Consequently, phonons folded from the $\Gamma$ point have degeneracy of one, while those from the X point, which has three equivalent wavevectors X, Y and Z, exhibit degeneracy of three. 
Table \ref{tab:mode-symmetry} lists the irreducible representations and degeneracies of each phonon branch \citep{RN371}. 
For the irreducible representations, Bouckaert-Smoluchowski-Wigner (BSW) notations are used \citep{RN366}. 
The phonon modes of the first branch is the soft phonon modes with imaginary frequency, which corresponds to cubic ($Fm\overline{3}m$) to tetragonal ($P4_2/nmc$) transition. 
The modes of the second branch are the acoustic modes, corresponding to translation of unitcell. 
The modes of the fourth branch are the only polar modes, which accounts for the spontaneous polarization in ferroelectric orthorhombic phase. 
The atom displacements of all the modes at $\Gamma$ point in the reciprocal space of cubic conventional unitcell are shown in the supplemental material. 
The polar modes with positive frequency imply that higher-order modes besides the soft mode are required to describe the ferroelectric transition in $\mathrm{HfO}_2$. 
We propose that we use the complete basis of all the modes to expand the atom displacements in unitcells of each phase. 
In the mode expansion of phases, the modes related to oxygen are crucial for ferroelectric properties, which are on the first, fourth, and sixth branches. 

\begin{figure}
    \includegraphics[scale=0.7]{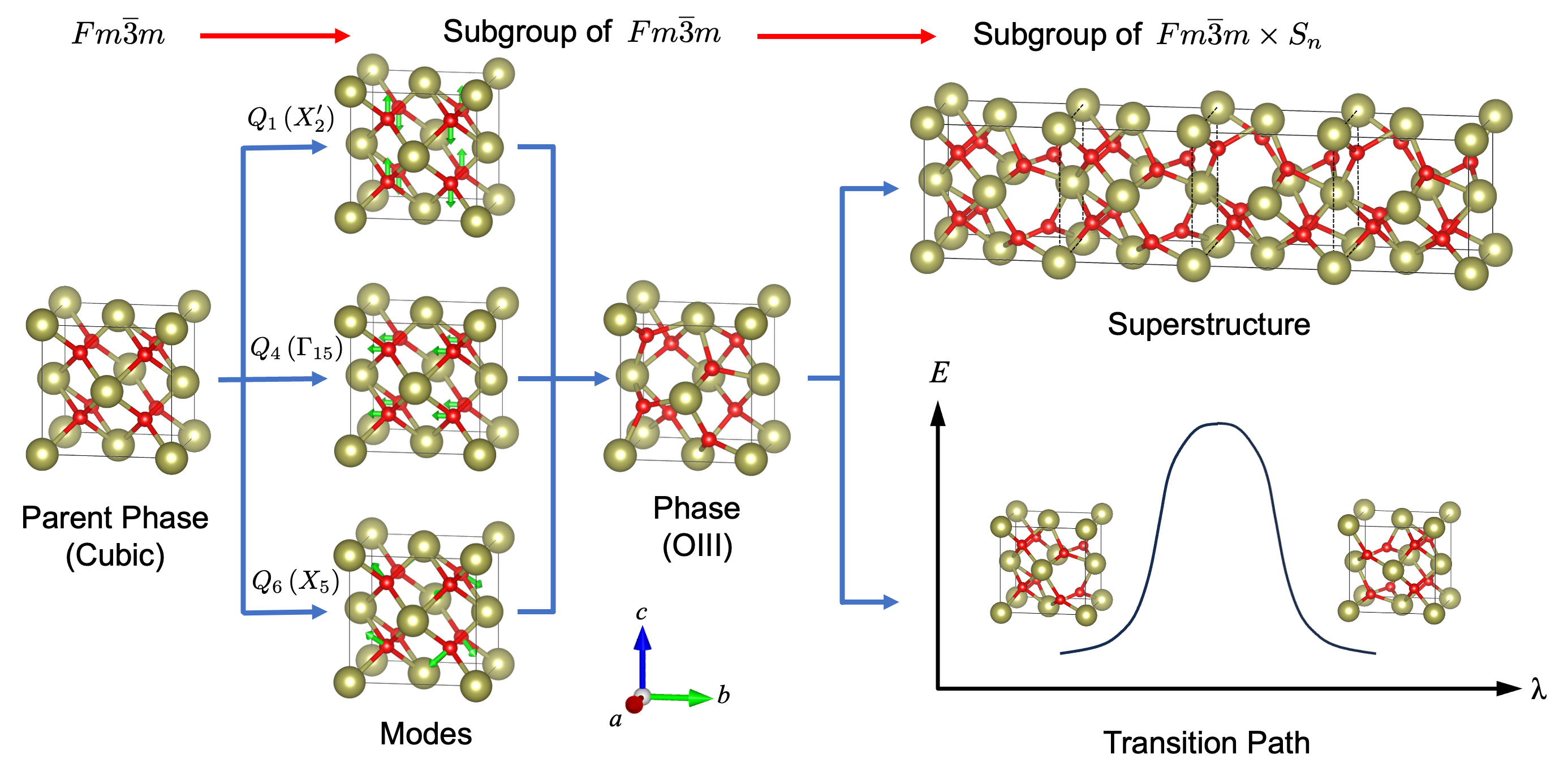}
    \caption{An unified framework to study ferroelectric properties of $\mathrm{HfO}_2$ using mode expansion. 
	The phases, superstructures such as domain walls and transition paths can be generated from cubic phase by adding phonon modes, and their symmetry groups are related to the symmetry group of parent cubic phase. 
    \label{fig:unified-framework}}
\end{figure}

\begin{table}[htbp]
    \centering
	\caption{Symmetries and degeneracies of phonon branches on $\Gamma$ point. Freq., dim., irrep. are the abbreviations of frequency, dimension and irreducible representation, respectively.\label{tab:mode-symmetry}}{%
    \begin{tabular}{c|cccccccc}
		Branch & Freq. (Hz) & Degeneracy & K point & Dim. of irrep. & Irrep. & Relevant element & Is polar? \\
        \hline
		1 & -6.38 & 3 & $X$      & 1 & $X_2^\prime$  & O    & No  \\
		2 & 0.0   & 3 & $\Gamma$ & 3 & $\Gamma_{15}$ & Hf,O & No  \\
		3 & 2.73  & 6 & $X$      & 2 & $X_5^\prime$  & Hf   & No  \\
		4 & 6.75  & 3 & $\Gamma$ & 3 & $\Gamma_{15}$ & O    & Yes \\
		5 & 7.53  & 3 & $X$      & 1 & $X_4^\prime$  & Hf   & No  \\
		6 & 9.06  & 6 & $X$      & 2 & $X_5$         & O    & No  \\
		7 & 16.8  & 6 & $X$      & 2 & $X_5^\prime$  & O    & No  \\
		8 & 17.3  & 3 & $\Gamma$ & 3 & $\Gamma_{25}^\prime$ & O    & No  \\
		9 & 20.7  & 3 & $X$      & 1 & $X_1$         & O    & No  \\
    \end{tabular}}{}
\end{table}

\section{Mode Expansion of Orthogonal System}

Next, we expand the atom displacements of orthogonal phases including cubic, tetragonal and ferroelectric orthorhombic phases by phonon modes. 
For the i-th branch with three-fold degeneracy, we use $Q_{ix}, Q_{iy}, Q_{iz}$ to denote the three modes of the branch.
Here the $x, y, z$ subscripts indicate the direction of atom displacements. 
For the j-th branch with six-fold degeneracy, we use $Q_{jzx}, Q_{jxy}, Q_{jyz}, Q_{jyx}, Q_{jzy}, Q_{jxz}$ to denote the six modes of the branch. 
Here the first subscripts X, Y and Z indicate the phonon mode is folded from X, Y and Z point in the primitive Brillouin zone, respectively. 
The second subscripts $x, y, z$ indicate the direction of atom displacements. 
We normalize these amplitudes so that the displacement of each atom along specific axis is the unit in relative coordinates. 
Equation \ref{eq:mode-expansion} shows the expansion of cubic (C), tetragonal (T), orthorhombic Pbcn and ferroelectric orthorhombic (OIII) phases by modes. 
It should be mentioned that the expansion is simplified. We neglects high-order modes and modes relevant to Hf atoms for better presentation, while retaining the symmetry properties under the cubic space group ($Fm\overline{3}m$). 
For the full expressions and the strain tensors, please refer to the supplemental material. 

\begin{equation}
	\begin{aligned}
		Q(\text{C}) &= 0 \\
		Q(\text{T}) &= 0.055 Q_{1z} \\
		Q(Pbcn) &= 0.11 Q_{1z} + 0.079 Q_{6xy} \\
		Q(\text{OIII}) &= 0.067 Q_{1z} + 0.054 Q_{4y} + 0.052 \left(Q_{6zx} + Q_{6xy} + Q_{6yz}\right)
	\end{aligned}
	\label{eq:mode-expansion}
\end{equation}

From the mode expansions, it becomes clear that the OIII phase is the only polar phase with polar mode $Q_{4y}$. 
The group-subgroup relations are satisfied: the space groups of OIII and Pbcn phases with more low-symmetry modes are the subgroups of T phase group and C phase group. 
Researchers have found that the unitcell of OIII phase with one of the Hf atoms fixed has 48 variants called pseudo-chirality structures when studying the transition path \citep{RN34,RN357,RN365}. 
These pseudo-chirality structures originate from the fact that the transformation group of OIII phase, $Pca2_1$, is the subgroup of cubic group $Fm\overline{3}m$. 
In the language of group theory \citep{RN370}, the symmetry operation on the crystal phase is the group action of space group G on the G-set, crystal phase. 
The symmetry transformations of OIII unitcell can be obtained by the transformations of modes (Fig. \ref{fig:unified-framework}). 
To find all the variants of OIII unitcell, we calculated the orbit of cubic group symmetry operation on one OIII unitcell, the set of unitcells linked by symmetry operation, using custom-developed Python package Hafnon. 
We found that the size of the orbit is 48, which means that there are 48 variants of OIII unitcells. 
For each OIII unitcell, we can apply mode expansion on it. The expansions shown in equation \ref{eq:expansion-OIII-variants} are similar in expression to the OIII expansion given above in equation \ref{eq:mode-expansion}, with some differences in signs and subscripts. 

\begin{equation}
	\begin{aligned}
		Q(\text{OIII,primary}) = Q_1\text{ term} + Q_4\text{ term} + 0.052 \left(\epsilon_1 Q_{6zx} + \epsilon_2 Q_{6xy} + \epsilon_3 Q_{6yz}\right) \\
		Q(\text{OIII,conjugate}) = Q_1\text{ term} + Q_4\text{ term} + 0.052 \left(\epsilon_1 Q_{6yx} + \epsilon_2 Q_{6zy} + \epsilon_3 Q_{6xz}\right)
	\end{aligned}
	\label{eq:expansion-OIII-variants}
\end{equation}

The dipole direction of OIII phase is related to modes of the fourth branch, which can only orient to one of the six crystal axes directions, $\pm a, \pm b, \pm c$. 
There are eight variants remaining if the dipole axis is fixed. The differences of these variants lead to the differences in the non-polar modes of the first and sixth branches. 
Because the modes of the first and fourth branches are related to the modes of the sixth branch, firstly, we focus on the the modes of the sixth branch. 
We found that only three of six modes of the sixth branch appear in the expansion, these triplet of active modes only have two combinations: $Q_{6zx}, Q_{6xy}, Q_{6yz}$ and $Q_{6yx}, Q_{6zy}, Q_{6xz}$. 
Under $C_4$ rotation ($90^\circ$ rotation), these two distinct triplets of modes transform to each other, forming a $C_4$-conjugate pair. 
We define the primary triplet as $Q_{6zx}, Q_{6xy}, Q_{6yz}$, and its conjugate counterpart as $Q_{6yx}, Q_{6zy}, Q_{6xz}$. 
The signs $\epsilon_1, \epsilon_2, \epsilon_3$ of these triplet modes must multiply to positive, leading to four distinct structures. 
The six dipole direction, the two distinct conjugate triplets and four sign combinations count for total 48 inequivalent structures. 
We use six crystal axis direction $\pm a, \pm b, \pm c$ to denote the dipole axis, and define a pseudo-chirality number ranged from 0 to 7 to 8 variants of unitcell with fixed dipole direction as in table \ref{tab:OIII-chirality-number}. 
The dipole axis and pseudo-chirality number uniquely determine all 48 variants of OIII unitcell. 
For compactness, we use overline for negative sign of axis, for instance, $\overline{c}$ stands for $-c$. 
For other phases, we can also assign a unique identifier for each variant of unitcell except for cubic phase which is invariant under symmetry transformations. 
The tetragonal phase has six variants which have different axis direction of the soft modes of the first branch, therefore, we use six crystal axis direction $\pm a, \pm b, \pm c$ to assign unique identifiers. 
The Pbcn phase has 24 variants. Though it has no polarization, we can also assign an axis direction and pseudo-chirality number ranged from 0 to 3, which is discussed in detail in the supplemental material. 

\begin{table}[htbp]
    \centering
	\caption{Signs and conjugate pairs in mode expansion of OIII and the definition of (pseudo-) chirality number \label{tab:OIII-chirality-number}}{%
    \begin{tabular}{ccccc}
		Chirality Number & Primary/Conjugate & $\epsilon_1$ & $\epsilon_2$ & $\epsilon_3$ \\
        \hline
		0 & P & + & + & + \\
		1 & P & + & − & − \\
		2 & P & − & + & − \\
		3 & P & − & − & − \\
		4 & C & + & + & + \\
		5 & C & + & − & − \\
		6 & C & − & + & − \\
		7 & C & − & − & + \\
    \end{tabular}}{}
\end{table}

\section{Energy Functional of Mode Amplitudes}

For the remaining two terms relevant to the modes of the first and fourth branches, the best way is to study the energy functional of modes and strain. 
Energy functional of ordering parameters has been studied recently \citep{RN364,RN349}. 
We expand the energy functional into polynomials of mode amplitudes and strain tensor. 
We focus on the terms relevant to the modes of first, fourth and sixth branches, which appear in mode expansion of OIII phase. 
The symmetry of the energy functional can reduce the number of terms significantly. 
Using custom-developed package Hafnon, we enumerated all the self-energy terms in the energy functional upto fourth-order terms which are invariant under symmetry transformations. 
The first-order terms in the energy functional are zero due to symmetry; the second-order terms cause the activation of modes in ferroelectric transition; and the fourth-order terms in the energy functional set the lower bound of energy, preventing the energy from dropping to negative infinity. 
The role of these terms is similar to the terms in the energy functional in Landau theory of ferroelectric transition, where the second-order term causes the instability of state at origin during phase transition, and the fourth-order term prevents the system from going to negative infinity. 
However, we found the existence of third-order terms in energy functional of $\mathrm{HfO}_2$, which is different from the case in $\mathrm{BaTiO}_3$ \citep{zhong1995first}. 
The third-order terms are found to be closely related to the stability of OIII phase \citep{RN349}. 
Furthermore, the existence of third-order terms indicates that the ferroelectric transition in $\mathrm{HfO}_2$ is likely to be a first-order transition, different from that in perovskite ferroelectrics. 
The diagonal part of the third-order terms, which is relevant to the modes of the same branch, is shown in equation \ref{eq:hamiltonian-3-order-diag}. 
Note that we omit a constant factor in the energy terms, which does not affect the physical conclusions. 
The off-diagonal part of the third-order terms, which contains products of modes of distinct branches, is shown in equation \ref{eq:hamiltonian-3-order-non-diag}. 
The activation of all the modes in the third-order terms minimizes the third-order energy functional, however, 
in most phases, fourth and higher-order terms suppress the activation of all the modes, leading to the minimization of the full energy functional. 
In the case of OIII phase, one and only one of the products in each third-order term is non-zero. 
Consequently, from the diagonal part in equation \ref{eq:hamiltonian-3-order-diag}, the activated modes of the sixth branch come in triplets of $Q_{6zx}, Q_{6xy}, Q_{6yz}$ or $Q_{6yx}, Q_{6zy}, Q_{6xz}$, and the signs multiply to positive. 
In addition, the two modes of the first and fourth branches, and one of the modes in the triplet should form a non-zero term in the off-diagonal part of the third order terms. 
The form of the off-diagonal part imposes restrictions on the two modes in the first and fourth branches: 
their displacement directions, together with the displacement direction of one triplet mode must be the complete permutation of x, y and z axes; 
their signs and the sign of the triplet mode multiply to positive. 
The selection of the mode in triplet to form a non-zero term is not arbitrary, in fact, only one triplet mode can be selected once the dipole axis, or the displacement axis of the mode of the fourth branch, is given. 
Based on energy functional analysis above, we summarize the rules to give a mode expansion of OIII unitcell with arbitrary dipole direction and pseudo-chirality number: 

\begin{enumerate}
	\item Write down the term relevant to modes of the sixth branch using the pseudo-chirality number table \ref{tab:OIII-chirality-number}. 
	\item Write down the term relevant to modes of the fourth branch based on dipole direction. We choose the mode with displacement axis same as the dipole axis, and write the same amplitude as the OIII expansion in equation \ref{eq:mode-expansion}. If the dipole direction is oriented along the negative axis, add a negative sign. 
	\item Write down the term relevant to modes of the first branch by analysing the off-diagonal part of third-order terms in equation \ref{eq:hamiltonian-3-order-non-diag}. Firstly, select one of the triplet to form a non-zero term in the off-diagonal terms with the modes of the first and fourth branches. The selected triplet mode should have different displacement axis with the polar mode (on the fourth branch). Then, we choose the mode from the first branch with displacement axis different from the mode of the fourth branch and sixth branch, set the amplitude to the same as that in equation \ref{eq:mode-expansion}, and assign sign to keep signs of these modes in the non-zero term multiply to positive. 
\end{enumerate}

\begin{equation}
	E_{3,diag} \propto -\left(Q_{6zx} Q_{6xy} Q_{6yz} + Q_{6yx} Q_{6zy} Q_{6xz}\right)
	\label{eq:hamiltonian-3-order-diag}
\end{equation}

\begin{equation}
	\begin{split}
		E_{3,non-diag} \propto -\left(Q_{1x} Q_{4y} Q_{6xz} + Q_{1x} Q_{4z} Q_{6xy} + Q_{1y} Q_{4x} Q_{6yz}\right. \\
		\left.+ Q_{1y} Q_{4z} Q_{6yx} + Q_{1z} Q_{4x} Q_{6zy} + Q_{1z} Q_{4y} Q_{6zx}\right)
	\end{split}
	\label{eq:hamiltonian-3-order-non-diag}
\end{equation}

Next, we analyze the strain terms in the energy functional. 
The strain tensor $e$ is second-order in the energy functional, as a consequence it has a linear term $e_{xx} + e_{yy} + e_{zz}$, also the only linear term in the energy functional. 
The next symmetrically-allowed term containing strain is the coupling of strain and phonon mode, which is first-order in strain, second-order in mode and fourth-order in energy functional. 
We focus on the terms containing shear strain shown in equation \ref{eq:hamiltonian-coupling} which plays a crucial role in the orthorhombic to monoclinic phase (M, $P2_1/c$) transition. 
The signs are positive, due to the meta-stability of OIII phase. 
In addition, these terms are zero in OIII phase because the dipole direction is along one of the crystal axes, and the activated modes of the sixth branch appear in triplets. 
During orthorhombic to monoclinic phase transition, these terms containing shear strain is small at the begining due to its fourth-order nature, which explains why the transition usually has a large barrier. 
This large barrier compared to the low barrier of orthorhombic to tetragonal transition leads to the kinetical stability of ferroelectric $\mathrm{HfO}_2$ \citep{RN256,RN134,RN160}, which prefers the orthorhombic to tetragonal pathway over orthorhombic to monoclinic pathway. 

\begin{equation}
	\begin{aligned}
		E_{coupling,1} \propto e_{xy} Q_{6zx} Q_{6zy} + e_{yz} Q_{6xy} Q_{6xz} + e_{zx} Q_{6yx} Q_{6yz} \\
		E_{coupling,2} \propto e_{xy} Q_{4x} Q_{4y} + e_{yz} Q_{4y} Q_{4z} + e_{zx} Q_{4x} Q_{4z}
	\end{aligned}
	\label{eq:hamiltonian-coupling}
\end{equation}

The energy functional analysis above shows that the stability of OIII phase has strong relation with the mode interactions and strain-mode coupling originated from the structure of energy surface. 
The full self-energy terms upto fourth-order are shown in supplemental material. 

\section{Ferroelectric Domain Wall}

Using mode expansion, we can build unitcells of orthogonal crystal phases and study their symmetry properties by the symmetry transformations of modes. 
For the superstructure consisting of unitcells that have aligned crystallographic orientations, we can also expand them into local modes that are localized in each unitcell (Fig. \ref{fig:unified-framework}). 
Note that they are not strictly aligned after structure relaxation due to the interactions between modes in adjacent cells. 
The transformation group of the superstructure consisting of $n$ unitcells is the subgroup of direct product of cubic space group and permutation group, $Fm\overline{3}m \times S_n$, in which the space group acts on local modes in each unitcell and the permutation group permutes $n$ unitcells. 

\begin{figure}
    \includegraphics[scale=0.7]{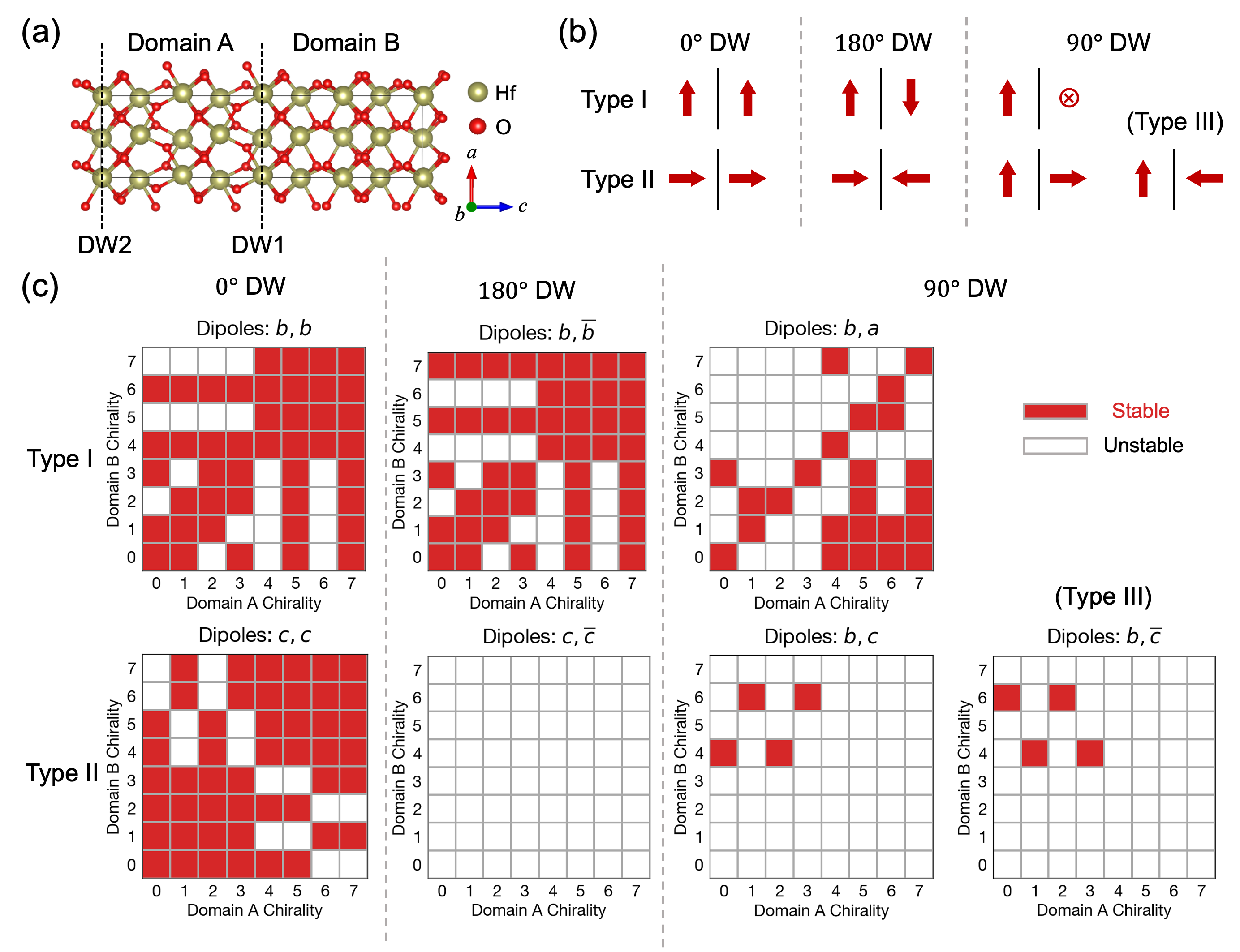}
    \caption{Effects of domain phonon modes on the stability of domain walls. 
	(a) Domain wall model comprised of two domains, A and B. 
	(b) The inequivalent dipole directions of two domains comprising the domain wall. 
	(c) Dependence of stability of domain walls on pseudo-chirality numbers of two domains, A and B. 
	The x-axis and y-axis of each stability map are the pseudo-chirality numbers of domain A and B in subfigure (a). 
	The figure title of each stability map indicates the dipole directions of domain A and B. 
    \label{fig:domain-wall-stability}}
\end{figure}

To study the effect of interface phonon modes on domain walls, we build a domain wall (DW) model consisting of two domains (Fig. \ref{fig:domain-wall-stability}a). 
The DW model is also a superstructure in Fig. \ref{fig:unified-framework} that can be expanded into local modes. 
Therefore, we can assign a unique identifier for each DW model using the sizes of two domains and the orthorhombic phase identifier of each domain, consisting of dipole direction and pseudo-chirality number. 
Besides DWs consisting of integer-size domains, the size of domain can be half-integer, because cubic phase is face-centered with translation in face diagonals and two half-integer sizes of two domains in our DW model add to integer. 
The total number of DW models is $48 \times 48 \times 2 = 4608$ for ferroelectric orthorhombic phase has 48 variants, which is costly for first-principle calculations. 
However, the symmetry of DW model can greatly reduce the number of DWs to calculate. 
The symmetry operation of DW model is the combination of cubic space group operation and permutation of two domains, in which the space group operation should not translate along the direction of normal vector of DW interface. 
The inequivalent DWs have been studied recently \citep{RN357,RN356}, in which they used hands with two colors to present the symmetry transformation of OIII unitcells and DWs. 
We proposed that the inequivalent DWs can be studied by the transformation of phonon modes which is more physical, and developed Hafnon package to count all the inequivalent DWs including the DW models with half-integer size domains. 
In group theory, the symmetry operation on DWs forms orbits, the sets of DWs linked by symmetry operation. 
The DWs belong to different orbits are inequivalent because no symmetry operation can transform them to each other. 
Using Hafnon package, we calculated the orbits of symmetry operation on DWs. 
These orbits divide the 2304 DWs into 134 equivalent classes, with one structure where no DW forms, 33 $0^\circ$ DWs, 34 $180^\circ$ DWs, and 66 $90^\circ$ DWs. 
After excluding the bad structures that have oxygen-oxygen bonds at interface, we have only 108 inequivalent good DWs, with one structure without DW, 23 good $0^\circ$ DWs, 30 good $180^\circ$ DWs, and 54 good $90^\circ$ DWs. 
We used first-principle calculations by VASP software \citep{RN109,RN110} to do structure relaxation of these inequivalent good DWs. 
Some DWs are not stable after structure relaxation, which can be classified into these cases: the disappearence of DW by DW motion; the transition from orthorhombic phase to monoclinic phase or tetragonal phase; the motion of DW by half-cell, increasing or decreasing domain size by half-integer; and the destruction of the whole superstructure. 
We found that the stability of DWs is related to the dipole directions and the pseudo-chirality number of two domains, which are directly related to interface phonon modes. 
There are total $6 \times 6 = 36$ dipole direction combinations of two domains, however, they are reduced to 7 types by calculating inequivalent classes, with 2 types of $0^\circ$ DWs, 2 types of $180^\circ$ DWs and 3 types of $90^\circ$ DWs. 
To determine the stability of arbitrary DWs, we transform them into DW with the standard dipole direction combinations, and look up the stability maps in Fig. \ref{fig:domain-wall-stability}c with the right dipole direction combination to find the stability of DWs with corresponding pseudo-chirality numbers. 
The relation of phonon modes and the DW stability and the effect of vacancy on DW stability are discussed in detail in our complementary work \bibnote{Please see our paper on arXiv entitled Domain Walls Stabilized by Intrinsic Phonon Modes and Engineered Defects Enable Robust Ferroelectricity in $\mathrm{HfO}_2$}. 

Using first-principle calculation, we found the stable condition of DW given the polarization directions and OIII pseudo-chirality number. 
Though the stable condition is summarized from the DW model with two domains (Fig. \ref{fig:domain-wall-stability}a), we can still explore the existence of ferroelectric topological domain such as ferroelectric vortex using these stability conditions. 
To form a ferroelectric vortex, the polarization vectors of distinct domains should form a loop \citep{RN372}. 
Given the polarization vectors, the stability of DWs is related to the pseudo-chirality number of each domain. 
The problem of finding the possible combinations of pseudo-chirality numbers is a variant of graph coloring problem in which the pseudo-chirality number act as colors, and the condition that no two adjacent vertices share the same color becomes the condition that the adjacent pseudo-chirality numbers stabilize the corresponding DW. 
We predicted several possible ferroelectric vortex structures in supplemental material. 

\section{Transition Path}

\begin{figure}
    \includegraphics[scale=0.8]{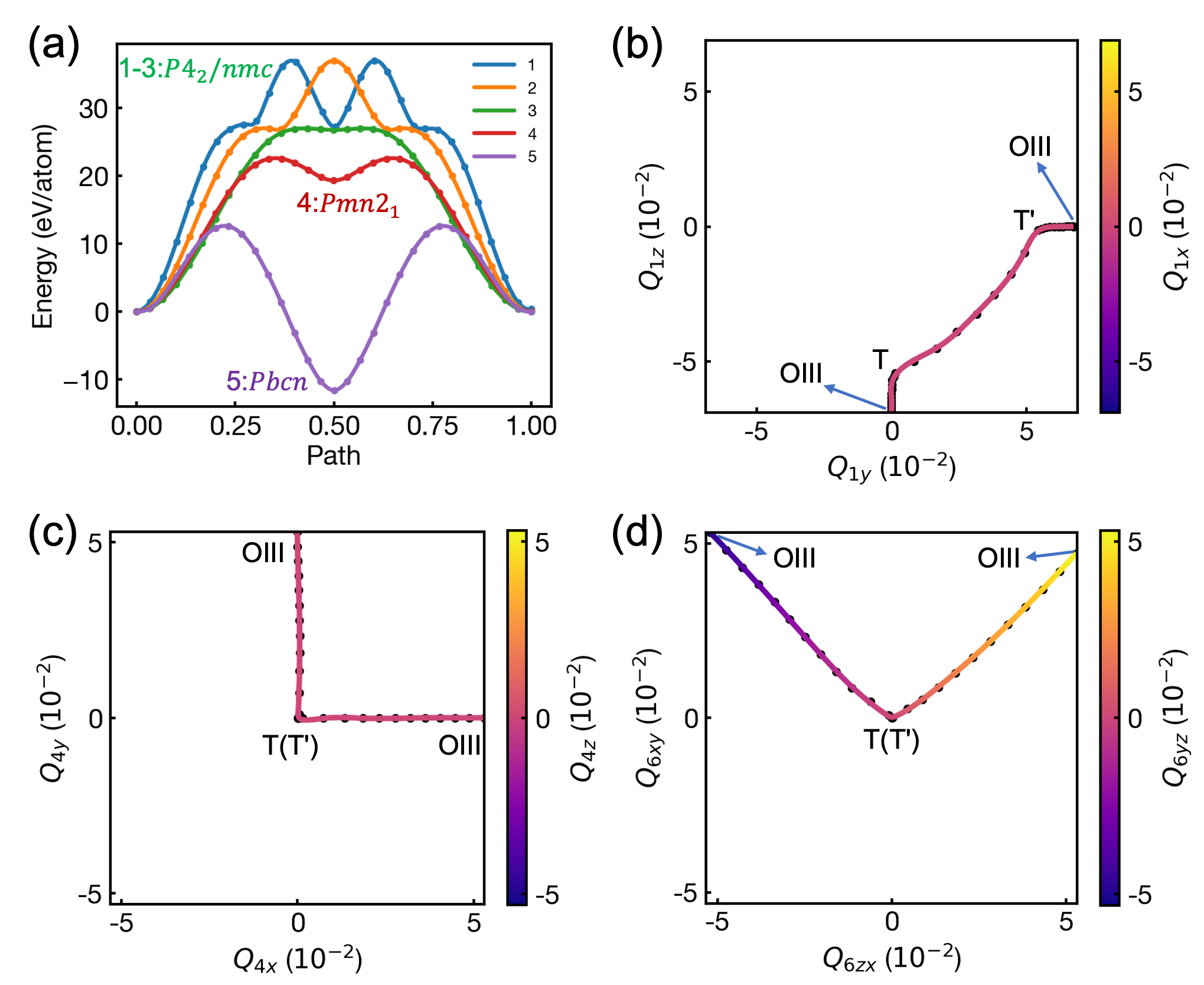}
    \caption{All possible switching paths of OIII phase and the evolution of modes along one of the paths. 
	(a) Five switching mechanisms of OIII phase. The IDs of switching paths are the same as these in Table \ref{tab:OIII-phase-switching}. 
	(b-d) The evolution of selected modes $Q_1$, $Q_4$ and the non-zero triplet of $Q_6$ along the $90^\circ$ transition path from OIII $a0$ unitcell to OIII $b2$ unitcell. The third dimension is visualized using colorbar. 
    \label{fig:OIII-phase-switching}}
\end{figure}

The transition path is of great importance in studying ferroelectric switching of ferroelectric materials. 
The widely acknowledged switching mechanism of OIII phase is the tetragonal mechanism in which tetragonal phase acts as the intermediate state of ferroelectric switching \citep{RN60}. 
Recently, researchers \citep{RN160,RN374} found that OIII phase has multiple transition paths originated from the variants of OIII unitcells, or pseudo-chirality. Some of the transition paths may involve the transition between tetragonal phase variants, which have larger barrier than paths without the transition between tetragonal variants. Qi \textit{et al.} found that whether the path requires switching of tetragonal phase is related to the continuity of $X_2^\prime$ modes of initial and final OIII phases \citep{RN349}. The paths with the same signs of $X_2^\prime$ modes of initial and final OIII phases have lower barrier than the paths with discontinued $X_2^\prime$ modes. 
Based on these facts, we can conclude that the phonon modes are important in the study of switching path. 
We used the same mode expansion analysis to study the inequivalent transition paths and the effect of modes on paths. 
The first step is to enumerate all the transition paths. 
We assumed that during switching the oxygen atoms do not move across Hf atomic planes, so that the path is unique for the same initial and final states. There exists paths where oxygen atoms move across the Hf atomic plane, but have larger barrier \citep{RN251}, therefore, we do not take them into consideration. 
Given the initial and final states, the path is unique, therefore, the number of transition paths is $48 \times 48 = 2304$, which is costly for firt principle calculation. 
Similarly, we used their symmetry group to reduce computational cost. 
The symmetry operation is simply the combination of cubic space group operation and permutation of initial and final states. 
The symmetry group orbits divide the 2304 paths into 18 equivalent classes, with one trivial path where the initial and final states are the same, four $0^\circ$ paths, five $180^\circ$ paths, and eight $90^\circ$ paths. 
Though there are 18 inequivalent paths, some of the paths have the same switching mechanism. 
We summarized the five switching mechanisms (there are six if we include the trivial path) in Table \ref{tab:OIII-phase-switching}, and plotted the energy as function of path by nudged-elastic band calculation in Fig. \ref{fig:OIII-phase-switching}. 
The five switching mechanisms fall in three categories, the T ($P4_2/nmc$) path, $Pmn2_1$ path and Pbcn path, in which there are three different T paths. We assign five IDs for the five switching mechanisms as in Fig. \ref{fig:OIII-phase-switching} and Table \ref{tab:OIII-phase-switching}. These five paths are in descending order of transition barriers. 
The third T path is the widely known ferroelectric switching path of OIII phase which has T phase as the intermediate state. 
The second T path can be divided into three parts: the first part is the transition from OIII phase to T phase; the second part is the transition from T phase unitcell to one of its variants, corresponding to the peak in the middle of the path; the third part is the transition from T phase to OIII phase. 
The first T path can also be divided into three parts, similar to the second T path, but the second part has two peaks. These two peaks corresponds to the transition from T phase unitcell to one of its variants, and another transition from the variant to the other variants of T unitcell. These peaks in the first and second paths raise the energy, increasing the transition barrier. 
The peaks in the middle parts of the first and second T paths are best explained when we expand the states on transition path by phonon modes. The transition path can be seen as a special phase whose mode amplitudes vary with path parameter. Instead of plotting each mode amplitude against the path parameter \citep{RN349}, we plotted the trajectory of mode amplitudes as a vector in 33-dimensional space in Fig. \ref{fig:OIII-phase-switching}b-d. The space is 33-dimensional because of the 36 degrees of freedom in unitcell with three translational degrees of freedom excluded. We chose three $Q_1$ modes, three $Q_4$ modes and the non-zero triplet of $Q_6$ modes to visualize in Fig. \ref{fig:OIII-phase-switching}b-d. One of the $90^\circ$ paths is chosen as an example to illustrate the origin of peaks. From Fig. \ref{fig:OIII-phase-switching}b, we can see that the norm of the $Q_1$ mode amplitude vector decreases first, corresponding to the transition from OIII phase to T phase. Then, the $Q_1$ mode amplitude vector rotates $90^\circ$, corresponding to the transition from T phase unitcell to one of its variants. Finally, the norm of rotated vector increases, corresponding to the transition from T phase to OIII phase. The $90^\circ$-rotation of $Q_1$ vector is due to the $90^\circ$ angle between the $Q_1$ modes of initial phase and final phases. When the angle between $Q_1$ modes of inital and final phases is $0^\circ$, the $Q_1$ mode will not rotate, instead it will increase to the value of OIII phase directly after its decrease. However, when the angle between $Q_1$ modes of inital and final phases is $180^\circ$, the $Q_1$ mode will not go straight to its opposite value. Instead, it decreases first, then rotates $180^\circ$ along an approximate half-circle, finally increases to the value of OIII phase. The reason why the $Q_1$ mode bypasses the origin of 33-dimensional space during transition is that there is a high energy peak at the origin which corresponds to the high-energy cubic phase. Therefore, to reduce the transition barrier, the $Q_1$ mode must be in the same direction for initial and final states like the third to fifth paths. As for the $Q_4$ mode and the non-zero triplet of $Q_6$ mode, they go almost straight from/to zero in the transition between OIII phase and T phase, while stay zero during the transition between variants of T phase. 
The fourth path has $Pmn2_1$ phase as the intermediate state, which only appears in $90^\circ$ switching. 
The last path, fifth path has Pbcn phase as the intermediate state, which only appears in $180^\circ$ switching. 
It should be noted that the phase energies of OIII and Pbcn phase are debatable, which are dependent on the choice of density functionals \citep{RN360}. 

\begin{table}[htbp]
    \centering
	\caption{Switching mechanism of OIII phase. Angle of dipoles is the angle between the dipoles of initial phase and final phase. Multiple values of angles mean that the angles of paths with corresponding mechanism can be one of these values. Percentage is the percentage of paths with corresponding switching mechanism in total $48 \times 48 = 2304$ paths. For paths equivalent to each other, only one of them is listed in the table. \label{tab:OIII-phase-switching}}{%
    \begin{tabular}{c|ccccc}
		ID & Mechanism & Angle of dipoles & Inequivalent paths & Percentage & Barrier (eV/atom) \\
        \hline
		1 & \makecell{$P4_2/nmc$ path \\ with two $90^\circ$ rotations \\ of $Q_1$ mode} & $0^\circ$, $90^\circ$, $180^\circ$ & \makecell{$c0 \to c1$, $c0 \to c3$, \\ $a4 \to b2$, $c0 \to \overline{c}0$, \\ $c0 \to \overline{c}2$}                                & 16.67\% & 37.0 \\
		2 & \makecell{$P4_2/nmc$ path \\ with $90^\circ$ rotation \\ of $Q_1$ mode}      & $0^\circ$, $90^\circ$, $180^\circ$ & \makecell{$c0 \to c4$, $a0 \to b0$, \\ $a0 \to b1$, $a0 \to b2$, \\ $a0 \to b3$, $a0 \to b4$, \\ $a0 \to b5$, $c0 \to \overline{c}4$} & 66.67\% & 37.0 \\
		3 & $P4_2/nmc$ path                                             & $0^\circ$, $180^\circ$             & $c0 \to c2$, $c0 \to \overline{c}1$                                                                               & 4.17\%  & 27.0 \\
		4 & $Pmn2_1$ path                                               & $90^\circ$                         & $a4 \to b0$                                                                                                       & 8.33\%  & 22.6 \\
		5 & $Pbcn$ path                                                 & $180^\circ$                        & $c0 \to \overline{c}3$                                                                                            & 2.08\%  & 12.6 \\
		6 & Trivial path (no transition)                                & $0^\circ$                          & $c0 \to c0$                                                                                                       & 2.08\%  & 0.0  \\
    \end{tabular}}{}
\end{table}

\section{Conclusion}

In summary, we developed a unified framework based on mode expansion for the studying of ferroelectricity in $\mathrm{HfO}_2$. We used cubic phase as the parent phase, expanding the atomic coordinates of orthogonal phases using a complete basis of phonon modes. 
To demonstrate the capability of this approach, we applied the mode expansion methods to the phase structures, DWs and transition paths of ferroelectric $\mathrm{HfO}_2$. 
We firstly calculated the phonon spectrum and phonon modes of C phase using first-principle calculations. 
After that, we made mode expansion of C, T, OIII and other orthogonal phases. 
By analysing the energy functional of mode amplitudes, we studied the stability of ferroelectric OIII phase and found that the modes play a crucial role in the stabilization of OIII phase. 
The domain walls are superstructures comprising of unitcells, which can also be expanded by the modes localized in unitcells. We enumerated all inequivalent domain walls and studied dependence of stability on interface modes by first-principle calculations. The stability of DW depends on the polarization directions and pseudo-chirality of domains, with some stability rules summarized in the article. 
Lastly, we summarized all switching mechanisms of OIII phase by enumerating inequivalent switching paths and calculating their barriers using nudged-elastic band method. By comparing the switching barriers of different paths, we found that the switching barrier is determined by the continuity of $Q_1$ modes and the polarization directions of initial and final states. 
This unified mode expansion approach can simplify the study of ferroelectric $\mathrm{HfO}_2$, especially the complicated superstructures including DWs and the transition paths. Also, this approach provides a new perspective of mode analysis for the exploration of ferroelectric properties. 

\pagebreak

\appendix*

\section{Methods}

Structure relaxation of DW model in Fig. \ref{fig:domain-wall-stability}a was carried out by first-principle calculation implemented in VASP software \citep{RN109,RN110}. 
Chemical species were treated using the projector-augmented plane wave method \citep{RN373}, with $5p^6 6s^2 5d^2$ valence electron configuration for Hf and $2s^2 2p^4$ for O. 
The plane wave cutoff energy was 600 eV and the Brillouin zone was sampled with gamma-centered k-point meshes with resolution 0.04. 
Exchange-correlation functional was approximated by the Perdew-Burke-Ernzerhof functional (PBE) \citep{RN148} in generalized gradiant approximation (GGA). 
The atom force convergence threshold was set to $0.01 \mathrm{eV}/\text{Å}$ and energy convergence threshold was set to $10^{-6} \mathrm{eV}$. 

DW model has one unitcell in a and b axis directions, and four unitcells in c axis direction, with two unitcells belonging to domain A and the other two belonging to domain B. 
The DW model has two domain wall interfaces due to three-dimensional periodic boundary condition as shown in Fig. \ref{fig:domain-wall-stability}a. 
All the DW structures are superstructures, which consist of unitcells that have approximately aligned crystallographic orientations, as shown in Fig. \ref{fig:domain-wall-stability}a. 
These superstructures were constructed from unitcells, which were themselves generated by mode expansion method, using Hafnon package developed by our group. 

The transition paths are constructed by linear interpolation of initial and final structures generated by by mode expansion method, and calculated by nudged-elastic band method implemented in VTST package \citep{RN147,RN146}. 
The path parameter is defined to be the distance between image and initial state, normalized so that the parameter value of final state is one. The distance of images is defined to be the minimum value of mean squared distance between atoms of images. 

The classification of inequivalent structures and paths is done by calculating group orbits implemented in Hafnon package. 

All the structures in figures were visualized using VESTA software \citep{RN189}. 

\begin{acknowledgments}
	This work is supported by the National Natural Science Foundation of China (Grants No. T2293703, No. T2293700).
\end{acknowledgments}

\bibliography{ref}

\clearpage
\widetext
\begin{center}
\textbf{\large Supplemental Materials for Decoding the Complexity of Ferroelectric Orthorhombic $\mathrm{HfO_2}$: A Unified Mode Expansion Approach}
\end{center}

\setcounter{section}{0}
\setcounter{page}{1}
\setcounter{secnumdepth}{2}
\appendix
\makeatletter

\renewcommand{\thefigure}{S\arabic{figure}}
\setcounter{figure}{0}

\renewcommand{\theequation}{S\arabic{equation}}
\setcounter{equation}{0}

\renewcommand{\thetable}{S\arabic{table}}
\setcounter{table}{0}

Supplemental material includes details of phonon modes, structure of cubic space group, the full expansions of orthogonal system phases, pseudo-chirality number of Pbcn phase, energy functional expanded upto fourth-order, and discussion on ferroelectric topological domains. 

\section{Phonon Modes}

Table \ref{tab:cubic-phase} shows the structure data of cubic phase. 

\begin{table}
    \caption{Cubic phase (conventional cell)\label{tab:cubic-phase}}{%
    \begin{tabular}{@{}cccc@{}}
        Formula & & $\mathrm{Hf}_4\mathrm{O}_8$ & \\
		\hline
        a, b, c (Å) & 5.07402 & 5.07402 & 5.07402 \\
        $\alpha$, $\beta$, $\gamma$ (deg) & 90.0000 & 90.0000 & 90.0000 \\
        \hline
        Atoms & & Relative coordinates & \\
		\hline
        Hf1 & 0.000000000 & 0.000000000 & 0.000000000 \\
        Hf2 & 0.000000000 & 0.500000000 & 0.500000000 \\
        Hf3 & 0.500000000 & 0.000000000 & 0.500000000 \\
        Hf4 & 0.500000000 & 0.500000000 & 0.000000000 \\
        O1 & 0.250000000 & 0.250000000 & 0.750000000 \\
        O2 & 0.250000000 & 0.750000000 & 0.750000000 \\
        O3 & 0.250000000 & 0.750000000 & 0.250000000 \\
        O4 & 0.250000000 & 0.250000000 & 0.250000000 \\
        O5 & 0.750000000 & 0.250000000 & 0.250000000 \\
        O6 & 0.750000000 & 0.750000000 & 0.250000000 \\
        O7 & 0.750000000 & 0.750000000 & 0.750000000 \\
        O8 & 0.750000000 & 0.250000000 & 0.750000000 \\
    \end{tabular}}{}
\end{table}

Tables \ref{tab:mode1-1} to \ref{tab:mode9-3} show the atom displacements of all the modes at $\Gamma$ point in the reciprocal space of cubic conventional unitcell. 
Disp. is the abbreviation for displacements. 

\begin{minipage}{\textwidth}
    \begin{minipage}{0.3\textwidth}
        \begin{table}[H]
            \begin{tabular}{cccc}
                \toprule
                Atoms & \multicolumn{3}{c}{Disp.} \\
                \midrule
                Hf1 & 0 & 0 & 0 \\
                Hf2 & 0 & 0 & 0 \\
                Hf3 & 0 & 0 & 0 \\
                Hf4 & 0 & 0 & 0 \\
                O5 & 1 & 0 & 0 \\
                O6 & -1 & 0 & 0 \\
                O7 & 1 & 0 & 0 \\
                O8 & -1 & 0 & 0 \\
                O9 & -1 & 0 & 0 \\
                O10 & 1 & 0 & 0 \\
                O11 & -1 & 0 & 0 \\
                O12 & 1 & 0 & 0 \\
                \bottomrule
            \end{tabular}
            \caption{Mode $Q_{1x}$\label{tab:mode1-1}}
        \end{table}
    \end{minipage}
    \begin{minipage}{0.3\textwidth}
        \begin{table}[H]
            \begin{tabular}{cccc}
                \toprule
                Atoms & \multicolumn{3}{c}{Disp.} \\
                \midrule
                Hf1 & 0 & 0 & 0 \\
                Hf2 & 0 & 0 & 0 \\
                Hf3 & 0 & 0 & 0 \\
                Hf4 & 0 & 0 & 0 \\
                O5 & 0 & 1 & 0 \\
                O6 & 0 & 1 & 0 \\
                O7 & 0 & -1 & 0 \\
                O8 & 0 & -1 & 0 \\
                O9 & 0 & 1 & 0 \\
                O10 & 0 & 1 & 0 \\
                O11 & 0 & -1 & 0 \\
                O12 & 0 & -1 & 0 \\
                \bottomrule
            \end{tabular}
            \caption{Mode $Q_{1y}$\label{tab:mode1-2}}
        \end{table}
    \end{minipage}
    \begin{minipage}{0.3\textwidth}
        \begin{table}[H]
            \begin{tabular}{cccc}
                \toprule
                Atoms & \multicolumn{3}{c}{Disp.} \\
                \midrule
                Hf1 & 0 & 0 & 0 \\
                Hf2 & 0 & 0 & 0 \\
                Hf3 & 0 & 0 & 0 \\
                Hf4 & 0 & 0 & 0 \\
                O5 & 0 & 0 & -1 \\
                O6 & 0 & 0 & 1 \\
                O7 & 0 & 0 & 1 \\
                O8 & 0 & 0 & -1 \\
                O9 & 0 & 0 & 1 \\
                O10 & 0 & 0 & -1 \\
                O11 & 0 & 0 & -1 \\
                O12 & 0 & 0 & 1 \\
                \bottomrule
            \end{tabular}
            \caption{Mode $Q_{1z}$\label{tab:mode1-3}}
        \end{table}
    \end{minipage}
\end{minipage}

\begin{minipage}{\textwidth}
    \begin{minipage}{0.3\textwidth}
        \begin{table}[H]
            \begin{tabular}{cccc}
                \toprule
                Atoms & \multicolumn{3}{c}{Disp.} \\
                \midrule
                Hf1 & 1 & 0 & 0 \\
                Hf2 & 1 & 0 & 0 \\
                Hf3 & 1 & 0 & 0 \\
                Hf4 & 1 & 0 & 0 \\
                O5 & 1 & 0 & 0 \\
                O6 & 1 & 0 & 0 \\
                O7 & 1 & 0 & 0 \\
                O8 & 1 & 0 & 0 \\
                O9 & 1 & 0 & 0 \\
                O10 & 1 & 0 & 0 \\
                O11 & 1 & 0 & 0 \\
                O12 & 1 & 0 & 0 \\
                \bottomrule
            \end{tabular}
            \caption{Mode $Q_{2x}$\label{tab:mode2-1}}
        \end{table}
    \end{minipage}
    \begin{minipage}{0.3\textwidth}
        \begin{table}[H]
            \begin{tabular}{cccc}
                \toprule
                Atoms & \multicolumn{3}{c}{Disp.} \\
                \midrule
                Hf1 & 0 & 1 & 0 \\
                Hf2 & 0 & 1 & 0 \\
                Hf3 & 0 & 1 & 0 \\
                Hf4 & 0 & 1 & 0 \\
                O5 & 0 & 1 & 0 \\
                O6 & 0 & 1 & 0 \\
                O7 & 0 & 1 & 0 \\
                O8 & 0 & 1 & 0 \\
                O9 & 0 & 1 & 0 \\
                O10 & 0 & 1 & 0 \\
                O11 & 0 & 1 & 0 \\
                O12 & 0 & 1 & 0 \\
                \bottomrule
            \end{tabular}
            \caption{Mode $Q_{2y}$\label{tab:mode2-2}}
        \end{table}
    \end{minipage}
    \begin{minipage}{0.3\textwidth}
        \begin{table}[H]
            \begin{tabular}{cccc}
                \toprule
                Atoms & \multicolumn{3}{c}{Disp.} \\
                \midrule
                Hf1 & 0 & 0 & 1 \\
                Hf2 & 0 & 0 & 1 \\
                Hf3 & 0 & 0 & 1 \\
                Hf4 & 0 & 0 & 1 \\
                O5 & 0 & 0 & 1 \\
                O6 & 0 & 0 & 1 \\
                O7 & 0 & 0 & 1 \\
                O8 & 0 & 0 & 1 \\
                O9 & 0 & 0 & 1 \\
                O10 & 0 & 0 & 1 \\
                O11 & 0 & 0 & 1 \\
                O12 & 0 & 0 & 1 \\
                \bottomrule
            \end{tabular}
            \caption{Mode $Q_{2z}$\label{tab:mode2-3}}
        \end{table}
    \end{minipage}
\end{minipage}

\begin{minipage}{\textwidth}
    \begin{minipage}{0.3\textwidth}
        \begin{table}[H]
            \begin{tabular}{cccc}
                \toprule
                Atoms & \multicolumn{3}{c}{Disp.} \\
                \midrule
                Hf1 & 1 & 0 & 0 \\
                Hf2 & -1 & 0 & 0 \\
                Hf3 & -1 & 0 & 0 \\
                Hf4 & 1 & 0 & 0 \\
                O5 & 0 & 0 & 0 \\
                O6 & 0 & 0 & 0 \\
                O7 & 0 & 0 & 0 \\
                O8 & 0 & 0 & 0 \\
                O9 & 0 & 0 & 0 \\
                O10 & 0 & 0 & 0 \\
                O11 & 0 & 0 & 0 \\
                O12 & 0 & 0 & 0 \\
                \bottomrule
            \end{tabular}
            \caption{Mode $Q_{3zx}$\label{tab:mode3-1}}
        \end{table}
    \end{minipage}
    \begin{minipage}{0.3\textwidth}
        \begin{table}[H]
            \begin{tabular}{cccc}
                \toprule
                Atoms & \multicolumn{3}{c}{Disp.} \\
                \midrule
                Hf1 & 0 & 1 & 0 \\
                Hf2 & 0 & 1 & 0 \\
                Hf3 & 0 & -1 & 0 \\
                Hf4 & 0 & -1 & 0 \\
                O5 & 0 & 0 & 0 \\
                O6 & 0 & 0 & 0 \\
                O7 & 0 & 0 & 0 \\
                O8 & 0 & 0 & 0 \\
                O9 & 0 & 0 & 0 \\
                O10 & 0 & 0 & 0 \\
                O11 & 0 & 0 & 0 \\
                O12 & 0 & 0 & 0 \\
                \bottomrule
            \end{tabular}
            \caption{Mode $Q_{3xy}$\label{tab:mode3-2}}
        \end{table}
    \end{minipage}
    \begin{minipage}{0.3\textwidth}
        \begin{table}[H]
            \begin{tabular}{cccc}
                \toprule
                Atoms & \multicolumn{3}{c}{Disp.} \\
                \midrule
                Hf1 & 0 & 0 & 1 \\
                Hf2 & 0 & 0 & -1 \\
                Hf3 & 0 & 0 & 1 \\
                Hf4 & 0 & 0 & -1 \\
                O5 & 0 & 0 & 0 \\
                O6 & 0 & 0 & 0 \\
                O7 & 0 & 0 & 0 \\
                O8 & 0 & 0 & 0 \\
                O9 & 0 & 0 & 0 \\
                O10 & 0 & 0 & 0 \\
                O11 & 0 & 0 & 0 \\
                O12 & 0 & 0 & 0 \\
                \bottomrule
            \end{tabular}
            \caption{Mode $Q_{3yz}$\label{tab:mode3-3}}
        \end{table}
    \end{minipage}
\end{minipage}

\begin{minipage}{\textwidth}
    \begin{minipage}{0.3\textwidth}
        \begin{table}[H]
            \begin{tabular}{cccc}
                \toprule
                Atoms & \multicolumn{3}{c}{Disp.} \\
                \midrule
                Hf1 & 1 & 0 & 0 \\
                Hf2 & -1 & 0 & 0 \\
                Hf3 & 1 & 0 & 0 \\
                Hf4 & -1 & 0 & 0 \\
                O5 & 0 & 0 & 0 \\
                O6 & 0 & 0 & 0 \\
                O7 & 0 & 0 & 0 \\
                O8 & 0 & 0 & 0 \\
                O9 & 0 & 0 & 0 \\
                O10 & 0 & 0 & 0 \\
                O11 & 0 & 0 & 0 \\
                O12 & 0 & 0 & 0 \\
                \bottomrule
            \end{tabular}
            \caption{Mode $Q_{3yx}$\label{tab:mode3-4}}
        \end{table}
    \end{minipage}
    \begin{minipage}{0.3\textwidth}
        \begin{table}[H]
            \begin{tabular}{cccc}
                \toprule
                Atoms & \multicolumn{3}{c}{Disp.} \\
                \midrule
                Hf1 & 0 & 1 & 0 \\
                Hf2 & 0 & -1 & 0 \\
                Hf3 & 0 & -1 & 0 \\
                Hf4 & 0 & 1 & 0 \\
                O5 & 0 & 0 & 0 \\
                O6 & 0 & 0 & 0 \\
                O7 & 0 & 0 & 0 \\
                O8 & 0 & 0 & 0 \\
                O9 & 0 & 0 & 0 \\
                O10 & 0 & 0 & 0 \\
                O11 & 0 & 0 & 0 \\
                O12 & 0 & 0 & 0 \\
                \bottomrule
            \end{tabular}
            \caption{Mode $Q_{3zy}$\label{tab:mode3-5}}
        \end{table}
    \end{minipage}
    \begin{minipage}{0.3\textwidth}
        \begin{table}[H]
            \begin{tabular}{cccc}
                \toprule
                Atoms & \multicolumn{3}{c}{Disp.} \\
                \midrule
                Hf1 & 0 & 0 & 1 \\
                Hf2 & 0 & 0 & 1 \\
                Hf3 & 0 & 0 & -1 \\
                Hf4 & 0 & 0 & -1 \\
                O5 & 0 & 0 & 0 \\
                O6 & 0 & 0 & 0 \\
                O7 & 0 & 0 & 0 \\
                O8 & 0 & 0 & 0 \\
                O9 & 0 & 0 & 0 \\
                O10 & 0 & 0 & 0 \\
                O11 & 0 & 0 & 0 \\
                O12 & 0 & 0 & 0 \\
                \bottomrule
            \end{tabular}
            \caption{Mode $Q_{3xz}$\label{tab:mode3-6}}
        \end{table}
    \end{minipage}
\end{minipage}

\begin{minipage}{\textwidth}
    \begin{minipage}{0.3\textwidth}
        \begin{table}[H]
            \begin{tabular}{cccc}
                \toprule
                Atoms & \multicolumn{3}{c}{Disp.} \\
                \midrule
                Hf1 & 0 & 0 & 0 \\
                Hf2 & 0 & 0 & 0 \\
                Hf3 & 0 & 0 & 0 \\
                Hf4 & 0 & 0 & 0 \\
                O5 & -1 & 0 & 0 \\
                O6 & -1 & 0 & 0 \\
                O7 & -1 & 0 & 0 \\
                O8 & -1 & 0 & 0 \\
                O9 & -1 & 0 & 0 \\
                O10 & -1 & 0 & 0 \\
                O11 & -1 & 0 & 0 \\
                O12 & -1 & 0 & 0 \\
                \bottomrule
            \end{tabular}
            \caption{Mode $Q_{4x}$\label{tab:mode4-1}}
        \end{table}
    \end{minipage}
    \begin{minipage}{0.3\textwidth}
        \begin{table}[H]
            \begin{tabular}{cccc}
                \toprule
                Atoms & \multicolumn{3}{c}{Disp.} \\
                \midrule
                Hf1 & 0 & 0 & 0 \\
                Hf2 & 0 & 0 & 0 \\
                Hf3 & 0 & 0 & 0 \\
                Hf4 & 0 & 0 & 0 \\
                O5 & 0 & -1 & 0 \\
                O6 & 0 & -1 & 0 \\
                O7 & 0 & -1 & 0 \\
                O8 & 0 & -1 & 0 \\
                O9 & 0 & -1 & 0 \\
                O10 & 0 & -1 & 0 \\
                O11 & 0 & -1 & 0 \\
                O12 & 0 & -1 & 0 \\
                \bottomrule
            \end{tabular}
            \caption{Mode $Q_{4y}$\label{tab:mode4-2}}
        \end{table}
    \end{minipage}
    \begin{minipage}{0.3\textwidth}
        \begin{table}[H]
            \begin{tabular}{cccc}
                \toprule
                Atoms & \multicolumn{3}{c}{Disp.} \\
                \midrule
                Hf1 & 0 & 0 & 0 \\
                Hf2 & 0 & 0 & 0 \\
                Hf3 & 0 & 0 & 0 \\
                Hf4 & 0 & 0 & 0 \\
                O5 & 0 & 0 & -1 \\
                O6 & 0 & 0 & -1 \\
                O7 & 0 & 0 & -1 \\
                O8 & 0 & 0 & -1 \\
                O9 & 0 & 0 & -1 \\
                O10 & 0 & 0 & -1 \\
                O11 & 0 & 0 & -1 \\
                O12 & 0 & 0 & -1 \\
                \bottomrule
            \end{tabular}
            \caption{Mode $Q_{4z}$\label{tab:mode4-3}}
        \end{table}
    \end{minipage}
\end{minipage}

\begin{minipage}{\textwidth}
    \begin{minipage}{0.3\textwidth}
        \begin{table}[H]
            \begin{tabular}{cccc}
                \toprule
                Atoms & \multicolumn{3}{c}{Disp.} \\
                \midrule
                Hf1 & 1 & 0 & 0 \\
                Hf2 & 1 & 0 & 0 \\
                Hf3 & -1 & 0 & 0 \\
                Hf4 & -1 & 0 & 0 \\
                O5 & 0 & 0 & 0 \\
                O6 & 0 & 0 & 0 \\
                O7 & 0 & 0 & 0 \\
                O8 & 0 & 0 & 0 \\
                O9 & 0 & 0 & 0 \\
                O10 & 0 & 0 & 0 \\
                O11 & 0 & 0 & 0 \\
                O12 & 0 & 0 & 0 \\
                \bottomrule
            \end{tabular}
            \caption{Mode $Q_{5x}$\label{tab:mode5-1}}
        \end{table}
    \end{minipage}
    \begin{minipage}{0.3\textwidth}
        \begin{table}[H]
            \begin{tabular}{cccc}
                \toprule
                Atoms & \multicolumn{3}{c}{Disp.} \\
                \midrule
                Hf1 & 0 & 1 & 0 \\
                Hf2 & 0 & -1 & 0 \\
                Hf3 & 0 & 1 & 0 \\
                Hf4 & 0 & -1 & 0 \\
                O5 & 0 & 0 & 0 \\
                O6 & 0 & 0 & 0 \\
                O7 & 0 & 0 & 0 \\
                O8 & 0 & 0 & 0 \\
                O9 & 0 & 0 & 0 \\
                O10 & 0 & 0 & 0 \\
                O11 & 0 & 0 & 0 \\
                O12 & 0 & 0 & 0 \\
                \bottomrule
            \end{tabular}
            \caption{Mode $Q_{5y}$\label{tab:mode5-2}}
        \end{table}
    \end{minipage}
    \begin{minipage}{0.3\textwidth}
        \begin{table}[H]
            \begin{tabular}{cccc}
                \toprule
                Atoms & \multicolumn{3}{c}{Disp.} \\
                \midrule
                Hf1 & 0 & 0 & 1 \\
                Hf2 & 0 & 0 & -1 \\
                Hf3 & 0 & 0 & -1 \\
                Hf4 & 0 & 0 & 1 \\
                O5 & 0 & 0 & 0 \\
                O6 & 0 & 0 & 0 \\
                O7 & 0 & 0 & 0 \\
                O8 & 0 & 0 & 0 \\
                O9 & 0 & 0 & 0 \\
                O10 & 0 & 0 & 0 \\
                O11 & 0 & 0 & 0 \\
                O12 & 0 & 0 & 0 \\
                \bottomrule
            \end{tabular}
            \caption{Mode $Q_{5z}$\label{tab:mode5-3}}
        \end{table}
    \end{minipage}
\end{minipage}

\begin{minipage}{\textwidth}
    \begin{minipage}{0.3\textwidth}
        \begin{table}[H]
            \begin{tabular}{cccc}
                \toprule
                Atoms & \multicolumn{3}{c}{Disp.} \\
                \midrule
                Hf1 & 0 & 0 & 0 \\
                Hf2 & 0 & 0 & 0 \\
                Hf3 & 0 & 0 & 0 \\
                Hf4 & 0 & 0 & 0 \\
                O5 & -1 & 0 & 0 \\
                O6 & -1 & 0 & 0 \\
                O7 & 1 & 0 & 0 \\
                O8 & 1 & 0 & 0 \\
                O9 & 1 & 0 & 0 \\
                O10 & 1 & 0 & 0 \\
                O11 & -1 & 0 & 0 \\
                O12 & -1 & 0 & 0 \\
                \bottomrule
            \end{tabular}
            \caption{Mode $Q_{6zx}$\label{tab:mode6-1}}
        \end{table}
    \end{minipage}
    \begin{minipage}{0.3\textwidth}
        \begin{table}[H]
            \begin{tabular}{cccc}
                \toprule
                Atoms & \multicolumn{3}{c}{Disp.} \\
                \midrule
                Hf1 & 0 & 0 & 0 \\
                Hf2 & 0 & 0 & 0 \\
                Hf3 & 0 & 0 & 0 \\
                Hf4 & 0 & 0 & 0 \\
                O5 & 0 & 1 & 0 \\
                O6 & 0 & 1 & 0 \\
                O7 & 0 & 1 & 0 \\
                O8 & 0 & 1 & 0 \\
                O9 & 0 & -1 & 0 \\
                O10 & 0 & -1 & 0 \\
                O11 & 0 & -1 & 0 \\
                O12 & 0 & -1 & 0 \\
                \bottomrule
            \end{tabular}
            \caption{Mode $Q_{6xy}$\label{tab:mode6-2}}
        \end{table}
    \end{minipage}
    \begin{minipage}{0.3\textwidth}
        \begin{table}[H]
            \begin{tabular}{cccc}
                \toprule
                Atoms & \multicolumn{3}{c}{Disp.} \\
                \midrule
                Hf1 & 0 & 0 & 0 \\
                Hf2 & 0 & 0 & 0 \\
                Hf3 & 0 & 0 & 0 \\
                Hf4 & 0 & 0 & 0 \\
                O5 & 0 & 0 & 1 \\
                O6 & 0 & 0 & -1 \\
                O7 & 0 & 0 & -1 \\
                O8 & 0 & 0 & 1 \\
                O9 & 0 & 0 & 1 \\
                O10 & 0 & 0 & -1 \\
                O11 & 0 & 0 & -1 \\
                O12 & 0 & 0 & 1 \\
                \bottomrule
            \end{tabular}
            \caption{Mode $Q_{6yz}$\label{tab:mode6-3}}
        \end{table}
    \end{minipage}
\end{minipage}

\begin{minipage}{\textwidth}
    \begin{minipage}{0.3\textwidth}
        \begin{table}[H]
            \begin{tabular}{cccc}
                \toprule
                Atoms & \multicolumn{3}{c}{Disp.} \\
                \midrule
                Hf1 & 0 & 0 & 0 \\
                Hf2 & 0 & 0 & 0 \\
                Hf3 & 0 & 0 & 0 \\
                Hf4 & 0 & 0 & 0 \\
                O5 & 1 & 0 & 0 \\
                O6 & -1 & 0 & 0 \\
                O7 & -1 & 0 & 0 \\
                O8 & 1 & 0 & 0 \\
                O9 & 1 & 0 & 0 \\
                O10 & -1 & 0 & 0 \\
                O11 & -1 & 0 & 0 \\
                O12 & 1 & 0 & 0 \\
                \bottomrule
            \end{tabular}
            \caption{Mode $Q_{6yx}$\label{tab:mode6-4}}
        \end{table}
    \end{minipage}
    \begin{minipage}{0.3\textwidth}
        \begin{table}[H]
            \begin{tabular}{cccc}
                \toprule
                Atoms & \multicolumn{3}{c}{Disp.} \\
                \midrule
                Hf1 & 0 & 0 & 0 \\
                Hf2 & 0 & 0 & 0 \\
                Hf3 & 0 & 0 & 0 \\
                Hf4 & 0 & 0 & 0 \\
                O5 & 0 & -1 & 0 \\
                O6 & 0 & -1 & 0 \\
                O7 & 0 & 1 & 0 \\
                O8 & 0 & 1 & 0 \\
                O9 & 0 & 1 & 0 \\
                O10 & 0 & 1 & 0 \\
                O11 & 0 & -1 & 0 \\
                O12 & 0 & -1 & 0 \\
                \bottomrule
            \end{tabular}
            \caption{Mode $Q_{6zy}$\label{tab:mode6-5}}
        \end{table}
    \end{minipage}
    \begin{minipage}{0.3\textwidth}
        \begin{table}[H]
            \begin{tabular}{cccc}
                \toprule
                Atoms & \multicolumn{3}{c}{Disp.} \\
                \midrule
                Hf1 & 0 & 0 & 0 \\
                Hf2 & 0 & 0 & 0 \\
                Hf3 & 0 & 0 & 0 \\
                Hf4 & 0 & 0 & 0 \\
                O5 & 0 & 0 & 1 \\
                O6 & 0 & 0 & 1 \\
                O7 & 0 & 0 & 1 \\
                O8 & 0 & 0 & 1 \\
                O9 & 0 & 0 & -1 \\
                O10 & 0 & 0 & -1 \\
                O11 & 0 & 0 & -1 \\
                O12 & 0 & 0 & -1 \\
                \bottomrule
            \end{tabular}
            \caption{Mode $Q_{6xz}$\label{tab:mode6-6}}
        \end{table}
    \end{minipage}
\end{minipage}

\begin{minipage}{\textwidth}
    \begin{minipage}{0.3\textwidth}
        \begin{table}[H]
            \begin{tabular}{cccc}
                \toprule
                Atoms & \multicolumn{3}{c}{Disp.} \\
                \midrule
                Hf1 & 0 & 0 & 0 \\
                Hf2 & 0 & 0 & 0 \\
                Hf3 & 0 & 0 & 0 \\
                Hf4 & 0 & 0 & 0 \\
                O5 & -1 & 0 & 0 \\
                O6 & 1 & 0 & 0 \\
                O7 & 1 & 0 & 0 \\
                O8 & -1 & 0 & 0 \\
                O9 & 1 & 0 & 0 \\
                O10 & -1 & 0 & 0 \\
                O11 & -1 & 0 & 0 \\
                O12 & 1 & 0 & 0 \\
                \bottomrule
            \end{tabular}
            \caption{Mode $Q_{7zx}$\label{tab:mode7-1}}
        \end{table}
    \end{minipage}
    \begin{minipage}{0.3\textwidth}
        \begin{table}[H]
            \begin{tabular}{cccc}
                \toprule
                Atoms & \multicolumn{3}{c}{Disp.} \\
                \midrule
                Hf1 & 0 & 0 & 0 \\
                Hf2 & 0 & 0 & 0 \\
                Hf3 & 0 & 0 & 0 \\
                Hf4 & 0 & 0 & 0 \\
                O5 & 0 & 1 & 0 \\
                O6 & 0 & -1 & 0 \\
                O7 & 0 & 1 & 0 \\
                O8 & 0 & -1 & 0 \\
                O9 & 0 & -1 & 0 \\
                O10 & 0 & 1 & 0 \\
                O11 & 0 & -1 & 0 \\
                O12 & 0 & 1 & 0 \\
                \bottomrule
            \end{tabular}
            \caption{Mode $Q_{7xy}$\label{tab:mode7-2}}
        \end{table}
    \end{minipage}
    \begin{minipage}{0.3\textwidth}
        \begin{table}[H]
            \begin{tabular}{cccc}
                \toprule
                Atoms & \multicolumn{3}{c}{Disp.} \\
                \midrule
                Hf1 & 0 & 0 & 0 \\
                Hf2 & 0 & 0 & 0 \\
                Hf3 & 0 & 0 & 0 \\
                Hf4 & 0 & 0 & 0 \\
                O5 & 0 & 0 & 1 \\
                O6 & 0 & 0 & 1 \\
                O7 & 0 & 0 & -1 \\
                O8 & 0 & 0 & -1 \\
                O9 & 0 & 0 & 1 \\
                O10 & 0 & 0 & 1 \\
                O11 & 0 & 0 & -1 \\
                O12 & 0 & 0 & -1 \\
                \bottomrule
            \end{tabular}
            \caption{Mode $Q_{7yz}$\label{tab:mode7-3}}
        \end{table}
    \end{minipage}
\end{minipage}

\begin{minipage}{\textwidth}
    \begin{minipage}{0.3\textwidth}
        \begin{table}[H]
            \begin{tabular}{cccc}
                \toprule
                Atoms & \multicolumn{3}{c}{Disp.} \\
                \midrule
                Hf1 & 0 & 0 & 0 \\
                Hf2 & 0 & 0 & 0 \\
                Hf3 & 0 & 0 & 0 \\
                Hf4 & 0 & 0 & 0 \\
                O5 & 1 & 0 & 0 \\
                O6 & 1 & 0 & 0 \\
                O7 & -1 & 0 & 0 \\
                O8 & -1 & 0 & 0 \\
                O9 & 1 & 0 & 0 \\
                O10 & 1 & 0 & 0 \\
                O11 & -1 & 0 & 0 \\
                O12 & -1 & 0 & 0 \\
                \bottomrule
            \end{tabular}
            \caption{Mode $Q_{7yx}$\label{tab:mode7-4}}
        \end{table}
    \end{minipage}
    \begin{minipage}{0.3\textwidth}
        \begin{table}[H]
            \begin{tabular}{cccc}
                \toprule
                Atoms & \multicolumn{3}{c}{Disp.} \\
                \midrule
                Hf1 & 0 & 0 & 0 \\
                Hf2 & 0 & 0 & 0 \\
                Hf3 & 0 & 0 & 0 \\
                Hf4 & 0 & 0 & 0 \\
                O5 & 0 & -1 & 0 \\
                O6 & 0 & 1 & 0 \\
                O7 & 0 & 1 & 0 \\
                O8 & 0 & -1 & 0 \\
                O9 & 0 & 1 & 0 \\
                O10 & 0 & -1 & 0 \\
                O11 & 0 & -1 & 0 \\
                O12 & 0 & 1 & 0 \\
                \bottomrule
            \end{tabular}
            \caption{Mode $Q_{7zy}$\label{tab:mode7-5}}
        \end{table}
    \end{minipage}
    \begin{minipage}{0.3\textwidth}
        \begin{table}[H]
            \begin{tabular}{cccc}
                \toprule
                Atoms & \multicolumn{3}{c}{Disp.} \\
                \midrule
                Hf1 & 0 & 0 & 0 \\
                Hf2 & 0 & 0 & 0 \\
                Hf3 & 0 & 0 & 0 \\
                Hf4 & 0 & 0 & 0 \\
                O5 & 0 & 0 & 1 \\
                O6 & 0 & 0 & -1 \\
                O7 & 0 & 0 & 1 \\
                O8 & 0 & 0 & -1 \\
                O9 & 0 & 0 & -1 \\
                O10 & 0 & 0 & 1 \\
                O11 & 0 & 0 & -1 \\
                O12 & 0 & 0 & 1 \\
                \bottomrule
            \end{tabular}
            \caption{Mode $Q_{7xz}$\label{tab:mode7-6}}
        \end{table}
    \end{minipage}
\end{minipage}

\begin{minipage}{\textwidth}
    \begin{minipage}{0.3\textwidth}
        \begin{table}[H]
            \begin{tabular}{cccc}
                \toprule
                Atoms & \multicolumn{3}{c}{Disp.} \\
                \midrule
                Hf1 & 0 & 0 & 0 \\
                Hf2 & 0 & 0 & 0 \\
                Hf3 & 0 & 0 & 0 \\
                Hf4 & 0 & 0 & 0 \\
                O5 & 1 & 0 & 0 \\
                O6 & -1 & 0 & 0 \\
                O7 & 1 & 0 & 0 \\
                O8 & -1 & 0 & 0 \\
                O9 & 1 & 0 & 0 \\
                O10 & -1 & 0 & 0 \\
                O11 & 1 & 0 & 0 \\
                O12 & -1 & 0 & 0 \\
                \bottomrule
            \end{tabular}
            \caption{Mode $Q_{8x}$\label{tab:mode8-1}}
        \end{table}
    \end{minipage}
    \begin{minipage}{0.3\textwidth}
        \begin{table}[H]
            \begin{tabular}{cccc}
                \toprule
                Atoms & \multicolumn{3}{c}{Disp.} \\
                \midrule
                Hf1 & 0 & 0 & 0 \\
                Hf2 & 0 & 0 & 0 \\
                Hf3 & 0 & 0 & 0 \\
                Hf4 & 0 & 0 & 0 \\
                O5 & 0 & 1 & 0 \\
                O6 & 0 & -1 & 0 \\
                O7 & 0 & 1 & 0 \\
                O8 & 0 & -1 & 0 \\
                O9 & 0 & 1 & 0 \\
                O10 & 0 & -1 & 0 \\
                O11 & 0 & 1 & 0 \\
                O12 & 0 & -1 & 0 \\
                \bottomrule
            \end{tabular}
            \caption{Mode $Q_{8y}$\label{tab:mode8-2}}
        \end{table}
    \end{minipage}
    \begin{minipage}{0.3\textwidth}
        \begin{table}[H]
            \begin{tabular}{cccc}
                \toprule
                Atoms & \multicolumn{3}{c}{Disp.} \\
                \midrule
                Hf1 & 0 & 0 & 0 \\
                Hf2 & 0 & 0 & 0 \\
                Hf3 & 0 & 0 & 0 \\
                Hf4 & 0 & 0 & 0 \\
                O5 & 0 & 0 & 1 \\
                O6 & 0 & 0 & -1 \\
                O7 & 0 & 0 & 1 \\
                O8 & 0 & 0 & -1 \\
                O9 & 0 & 0 & 1 \\
                O10 & 0 & 0 & -1 \\
                O11 & 0 & 0 & 1 \\
                O12 & 0 & 0 & -1 \\
                \bottomrule
            \end{tabular}
            \caption{Mode $Q_{8z}$\label{tab:mode8-3}}
        \end{table}
    \end{minipage}
\end{minipage}

\begin{minipage}{\textwidth}
    \begin{minipage}{0.3\textwidth}
        \begin{table}[H]
            \begin{tabular}{cccc}
                \toprule
                Atoms & \multicolumn{3}{c}{Disp.} \\
                \midrule
                Hf1 & 0 & 0 & 0 \\
                Hf2 & 0 & 0 & 0 \\
                Hf3 & 0 & 0 & 0 \\
                Hf4 & 0 & 0 & 0 \\
                O5 & -1 & 0 & 0 \\
                O6 & -1 & 0 & 0 \\
                O7 & -1 & 0 & 0 \\
                O8 & -1 & 0 & 0 \\
                O9 & 1 & 0 & 0 \\
                O10 & 1 & 0 & 0 \\
                O11 & 1 & 0 & 0 \\
                O12 & 1 & 0 & 0 \\
                \bottomrule
            \end{tabular}
            \caption{Mode $Q_{9x}$\label{tab:mode9-1}}
        \end{table}
    \end{minipage}
    \begin{minipage}{0.3\textwidth}
        \begin{table}[H]
            \begin{tabular}{cccc}
                \toprule
                Atoms & \multicolumn{3}{c}{Disp.} \\
                \midrule
                Hf1 & 0 & 0 & 0 \\
                Hf2 & 0 & 0 & 0 \\
                Hf3 & 0 & 0 & 0 \\
                Hf4 & 0 & 0 & 0 \\
                O5 & 0 & 1 & 0 \\
                O6 & 0 & -1 & 0 \\
                O7 & 0 & -1 & 0 \\
                O8 & 0 & 1 & 0 \\
                O9 & 0 & 1 & 0 \\
                O10 & 0 & -1 & 0 \\
                O11 & 0 & -1 & 0 \\
                O12 & 0 & 1 & 0 \\
                \bottomrule
            \end{tabular}
            \caption{Mode $Q_{9y}$\label{tab:mode9-2}}
        \end{table}
    \end{minipage}
    \begin{minipage}{0.3\textwidth}
        \begin{table}[H]
            \begin{tabular}{cccc}
                \toprule
                Atoms & \multicolumn{3}{c}{Disp.} \\
                \midrule
                Hf1 & 0 & 0 & 0 \\
                Hf2 & 0 & 0 & 0 \\
                Hf3 & 0 & 0 & 0 \\
                Hf4 & 0 & 0 & 0 \\
                O5 & 0 & 0 & 1 \\
                O6 & 0 & 0 & 1 \\
                O7 & 0 & 0 & -1 \\
                O8 & 0 & 0 & -1 \\
                O9 & 0 & 0 & -1 \\
                O10 & 0 & 0 & -1 \\
                O11 & 0 & 0 & 1 \\
                O12 & 0 & 0 & 1 \\
                \bottomrule
            \end{tabular}
            \caption{Mode $Q_{9z}$\label{tab:mode9-3}}
        \end{table}
    \end{minipage}
\end{minipage}

\section{Cubic Space Group}

The cubic space group $Fm\overline{3}m$ is the semidirect product of a translation group and a rotation group, $G_T \rtimes G_R$. 
In terms of cubic space group, we refer to the group that transforms the conventional unitcell of cubic phase. 

Translation group elements are (the three components can add arbitrary integers): 

\begin{equation}
    G_T = 
    \begin{pmatrix}
    0 \\
    0 \\
    0
    \end{pmatrix},
    \begin{pmatrix}
    0 \\
    1/2 \\
    1/2
    \end{pmatrix},
    \begin{pmatrix}
    1/2 \\
    0 \\
    1/2
    \end{pmatrix},
    \begin{pmatrix}
    1/2 \\
    1/2 \\
    0
    \end{pmatrix}
\end{equation}

Rotation group elements are: 

\begin{equation}
    \begin{aligned}
    G_R =& 
    \begin{pmatrix}
    \pm 1 & 0 & 0 \\
    0 & \pm 1 & 0 \\
    0 & 0 & \pm 1
    \end{pmatrix},
    \begin{pmatrix}
    0 & 0 & \pm 1 \\
    0 & \pm 1 & 0 \\
    \pm 1 & 0 & 0
    \end{pmatrix},
    \begin{pmatrix}
    \pm 1 & 0 & 0 \\
    0 & 0 & \pm 1 \\
    0 & \pm 1 & 0
    \end{pmatrix}, \\
    &\begin{pmatrix}
    0 & \pm 1 & 0 \\
    \pm 1 & 0 & 0 \\
    0 & 0 & \pm 1
    \end{pmatrix},
    \begin{pmatrix}
    0 & 0 & \pm 1 \\
    \pm 1 & 0 & 0 \\
    0 & \pm 1 & 0
    \end{pmatrix},
    \begin{pmatrix}
    0 & \pm 1 & 0 \\
    0 & 0 & \pm 1 \\
    \pm 1 & 0 & 0
    \end{pmatrix}
    \end{aligned}
\end{equation}

Group action on fractional coordinates $x$ is defined to be:

\begin{equation}
    (t, R) \in G_T \rtimes G_R,\, (t, R)x = Rx + t
\end{equation}

Group multiplication is defined to be:

\begin{equation}
    (t_1, R_1) \cdot (t_2, R_2) = (R_1 t_2 + t_1, R_1 R_2)
\end{equation}

\section{Decomposition of Cubic Space Group Action}

The group actions on phases, domain walls or transtion paths are simplified after decomposition into \textit{elementary} actions. The decomposition of group actions can be done by the decomposition of group elements. 
According to the discussion in the previous section, we can decompose the cubic group elements into the multiplication of a translation ($t_x, t_y, t_z$) and a point group element. The point group element can be further decomposed into the multiplication of three mirroring elements ($\sigma_x, \sigma_y, \sigma_z$), the $C_{4y}$ rotation and the $C_{4z}$ rotation:

\begin{equation}
    g = t_x t_y t_z \sigma_x \sigma_y \sigma_z C_{4y} C_{4z}
    \label{eq:group-action-decomp}
\end{equation}

Note that all of the elements in the decomposition of cubic group elements (Equation \ref{eq:group-action-decomp}) can be substituted by the identity element. This decomposition expression can be easily written if we known that we can always transform the point group element into diagonal form after multiplication of $C_{4y}$ and $C_{4z}$. 
After the decomposition of cubic group element, we do the group action by doing each elementary group action from the right of decomposition expression to the left. This implementation of group action reduces the number of group actions to be implemented to eight, and offers a unified way to treat all the group actions such as unitcells, domain walls and transition paths. 

\section{Full Expansion of Orthogonal System Phases}

Expansions of cubic phase ($Fm\overline{3}m$) and tetragonal phase ($P4_2/nmc$) shown in the main article are already full expansions. 

Full expansion of Pbcn phase

\begin{equation}
	Q(Pbcn) = 0.11 Q_{1z} + 0.075 Q_{3yz} + 0.079 Q_{6xy} + 0.025 Q_{7yx}
\end{equation}

Full expansion of ferroelectric orthorhombic phase ($Pca2_1$)

\begin{equation}
    \begin{aligned}
        Q(\text{OIII}) &= 0.067 Q_{1z} + 0.033 Q_{3yz} + 0.054 Q_{4y} + 0.017 Q_{5x} \\
        & + 0.053 Q_{6zx} + 0.054 Q_{6xy} + 0.048 Q_{6yz} + 0.015 Q_{7yx}
    \end{aligned}
\end{equation}

\section{Chirality Number of Pbcn Phase}

The expansion of simplified Pbcn phase is shown below: 

\begin{equation}
    Q(Pbcn) = 0.11 Q_1 + 0.079 Q_6
\end{equation}

We assign the unique identifier of Pbcn phase to be one axis direction ($\pm a, \pm b, \pm c$) and pseudo-chirality number (total four numbers). 
The dependence of terms $Q_1, Q_6$ to axis $d$ and pseudo-chirality number is as follows: 

\begin{table}[htbp]
    \centering
	\caption{Definition of Pbcn phase (pseudo-) chirality number}
    \begin{tabular}{cccc}
		Axis & Chirality Number & $Q_1$ & $Q_6$ \\
        \hline
		$d$ & 0 & $Q_{1s_2}$ & $Q_{6s_3s_1}$ \\
		$d$ & 1 & $-Q_{1s_2}$ & $-Q_{6s_3s_1}$ \\
		$d$ & 2 & $Q_{1s_3}$ & $Q_{6s_2s_1}$ \\
		$d$ & 3 & $-Q_{1s_3}$ & $-Q_{6s_2s_1}$ \\
        $-d$ & 0 & $-Q_{1s_2}$ & $Q_{6s_3s_1}$ \\
		$-d$ & 1 & $Q_{1s_2}$ & $-Q_{6s_3s_1}$ \\
		$-d$ & 2 & $-Q_{1s_3}$ & $Q_{6s_2s_1}$ \\
		$-d$ & 3 & $Q_{1s_3}$ & $-Q_{6s_2s_1}$ \\
    \end{tabular}
\end{table}

Where the relation between axis $d$ and subscripts $s_1, s_2, s_3$ is as follows: 

\begin{table}[htbp]
    \centering
	\caption{Relation between axis $d$ and subscripts $s_1, s_2, s_3$}{%
    \begin{tabular}{cccc}
		Axis ($d$) & $s_1$ & $s_2$ & $s_3$ \\
        \hline
		$a$ & $x$ & $y$ & $z$ \\
		$b$ & $y$ & $z$ & $x$ \\
		$c$ & $z$ & $x$ & $y$ \\
    \end{tabular}}{}
\end{table}

\section{Energy Functional Expanded upto Fourth-order}

\subsection{Second-Order Mode Terms}

\begin{equation}
	\begin{aligned}
		E(Q_1^2) &= Q_{1x}^2 + Q_{1y}^2 + Q_{1z}^2 \\
		E(Q_4^2) &= Q_{4x}^2 + Q_{4y}^2 + Q_{4z}^2 \\
		E(Q_6^2) &= Q_{6xy}^2 + Q_{6xz}^2 + Q_{6yx}^2 + Q_{6yz}^2 + Q_{6zx}^2 + Q_{6zy}^2
	\end{aligned}
\end{equation}

\subsection{Third-Order Mode Terms}

\begin{equation}
	\begin{aligned}
		E(Q_6^3) &= Q_{6xy} Q_{6yz} Q_{6zx} + Q_{6xz} Q_{6yx} Q_{6zy} \\
		E(Q_1 Q_4 Q_6) &= Q_{1x} Q_{4y} Q_{6xz} + Q_{1x} Q_{4z} Q_{6xy} + Q_{1y} Q_{4x} Q_{6yz} \\
        &+ Q_{1y} Q_{4z} Q_{6yx} + Q_{1z} Q_{4x} Q_{6zy} + Q_{1z} Q_{4y} Q_{6zx}
	\end{aligned}
\end{equation}

\subsection{Fourth-Order Mode Terms, Diagonal Part}

Terms relevant to $Q_1$

\begin{equation}
	\begin{aligned}
		E_1(Q_1^4) &= Q_{1x}^4 + Q_{1y}^4 + Q_{1z}^4 \\
		E_2(Q_1^4) &= Q_{1x}^2 Q_{1y}^2 + Q_{1x}^2 Q_{1z}^2 + Q_{1y}^2 Q_{1z}^2
	\end{aligned}
\end{equation}

Terms relevant to $Q_4$

\begin{equation}
	\begin{aligned}
		E_1(Q_4^4) &= Q_{4x}^4 + Q_{4y}^4 + Q_{4z}^4 \\
		E_2(Q_4^4) &= Q_{4x}^2 Q_{4y}^2 + Q_{4x}^2 Q_{4z}^2 + Q_{4y}^2 Q_{4z}^2
	\end{aligned}
\end{equation}

Terms relevant to $Q_6$

\begin{equation}
	\begin{aligned}
		E_1(Q_6^4) &= Q_{6xy}^4 + Q_{6xz}^4 + Q_{6yx}^4 + Q_{6yz}^4 + Q_{6zx}^4 + Q_{6zy}^4 \\
		E_2(Q_6^4) &= Q_{6xy}^2 Q_{6yz}^2 + Q_{6xy}^2 Q_{6zx}^2 + Q_{6xz}^2 Q_{6yx}^2 \\
        &+ Q_{6xz}^2 Q_{6zy}^2 + Q_{6yx}^2 Q_{6zy}^2 + Q_{6yz}^2 Q_{6zx}^2 \\
		E_3(Q_6^4) &= Q_{6xy}^2 Q_{6zy}^2 + Q_{6xz}^2 Q_{6yz}^2 + Q_{6yx}^2 Q_{6zx}^2 \\
		E_4(Q_6^4) &= Q_{6xy}^2 Q_{6yx}^2 + Q_{6xz}^2 Q_{6zx}^2 + Q_{6yz}^2 Q_{6zy}^2 \\
		E_5(Q_6^4) &= Q_{6xy}^2 Q_{6xz}^2 + Q_{6yx}^2 Q_{6yz}^2 + Q_{6zx}^2 Q_{6zy}^2
	\end{aligned}
\end{equation}

\subsection{Fourth-Order Mode Terms, Off-diagonal Part}

Coupling of $Q_1$ and $Q_4$

\begin{equation}
	\begin{aligned}
		E_1(Q_1^2 Q_4^2) &= Q_{1x}^2 Q_{4x}^2 + Q_{1y}^2 Q_{4y}^2 + Q_{1z}^2 Q_{4z}^2 \\
		E_2(Q_1^2 Q_4^2) &= Q_{1x}^2 Q_{4y}^2 + Q_{1x}^2 Q_{4z}^2 + Q_{1y}^2 Q_{4x}^2 + Q_{1y}^2 Q_{4z}^2 + Q_{1z}^2 Q_{4x}^2 + Q_{1z}^2 Q_{4y}^2
	\end{aligned}
\end{equation}

Coupling of $Q_1$ and $Q_6$

\begin{equation}
	\begin{aligned}
		E_1(Q_1^2 Q_6^2) &= Q_{1x} Q_{1y} Q_{6xy} Q_{6yx} + Q_{1x} Q_{1z} Q_{6xz} Q_{6zx} + Q_{1y} Q_{1z} Q_{6yz} Q_{6zy} \\
		E_2(Q_1^2 Q_6^2) &= Q_{1x}^2 Q_{6yx}^2 + Q_{1x}^2 Q_{6zx}^2 + Q_{1y}^2 Q_{6xy}^2 + Q_{1y}^2 Q_{6zy}^2 + Q_{1z}^2 Q_{6xz}^2 + Q_{1z}^2 Q_{6yz}^2 \\
		E_3(Q_1^2 Q_6^2) &= Q_{1x}^2 Q_{6xy}^2 + Q_{1x}^2 Q_{6xz}^2 + Q_{1y}^2 Q_{6yx}^2 + Q_{1y}^2 Q_{6yz}^2 + Q_{1z}^2 Q_{6zx}^2 + Q_{1z}^2 Q_{6zy}^2 \\
		E_4(Q_1^2 Q_6^2) &= Q_{1x}^2 Q_{6yz}^2 + Q_{1x}^2 Q_{6zy}^2 + Q_{1y}^2 Q_{6xz}^2 + Q_{1y}^2 Q_{6zx}^2 + Q_{1z}^2 Q_{6xy}^2 + Q_{1z}^2 Q_{6yx}^2
	\end{aligned}
\end{equation}

Coupling of $Q_4$ and $Q_6$

\begin{equation}
	\begin{aligned}
		E_1(Q_4^2 Q_6^2) &= Q_{4x} Q_{4y} Q_{6zx} Q_{6zy} + Q_{4x} Q_{4z} Q_{6yx} Q_{6yz} + Q_{4y} Q_{4z} Q_{6xy} Q_{6xz} \\
		E_2(Q_4^2 Q_6^2) &= Q_{4x}^2 Q_{6yx}^2 + Q_{4x}^2 Q_{6zx}^2 + Q_{4y}^2 Q_{6xy}^2 + Q_{4y}^2 Q_{6zy}^2 + Q_{4z}^2 Q_{6xz}^2 + Q_{4z}^2 Q_{6yz}^2 \\
		E_3(Q_4^2 Q_6^2) &= Q_{4x}^2 Q_{6xy}^2 + Q_{4x}^2 Q_{6xz}^2 + Q_{4y}^2 Q_{6yx}^2 + Q_{4y}^2 Q_{6yz}^2 + Q_{4z}^2 Q_{6zx}^2 + Q_{4z}^2 Q_{6zy}^2 \\
		E_4(Q_4^2 Q_6^2) &= Q_{4x}^2 Q_{6yz}^2 + Q_{4x}^2 Q_{6zy}^2 + Q_{4y}^2 Q_{6xz}^2 + Q_{4y}^2 Q_{6zx}^2 + Q_{4z}^2 Q_{6xy}^2 + Q_{4z}^2 Q_{6yx}^2
	\end{aligned}
\end{equation}

Coupling of $Q_1$, $Q_4$ and $Q_6$

\begin{equation}
    \begin{aligned}
        E_1(Q_1 Q_4 Q_6^2) &= Q_{1x} Q_{4x} Q_{6yx} Q_{6zx} + Q_{1y} Q_{4y} Q_{6xy} Q_{6zy} + Q_{1z} Q_{4z} Q_{6xz} Q_{6yz} \\
        E_2(Q_1 Q_4 Q_6^2) &= Q_{1x} Q_{4y} Q_{6yx} Q_{6zy} + Q_{1x} Q_{4z} Q_{6yz} Q_{6zx} + Q_{1y} Q_{4x} Q_{6xy} Q_{6zx} \\
        &+ Q_{1y} Q_{4z} Q_{6xz} Q_{6zy} + Q_{1z} Q_{4x} Q_{6xz} Q_{6yx} + Q_{1z} Q_{4y} Q_{6xy} Q_{6yz}
    \end{aligned}
\end{equation}

\subsection{First-Order Strain Term}

\begin{equation}
	E(e) = e_{xx} + e_{yy} + e_{zz}
\end{equation}

\subsection{Second-Order Strain Terms}

\begin{equation}
	\begin{aligned}
		E_1(e^2) &= e_{xx}^2 + e_{yy}^2 + e_{zz}^2 \\
		E_2(e^2) &= e_{xy}^2 + e_{yz}^2 + e_{zx}^2 \\
		E_3(e^2) &= e_{xx} e_{yy} + e_{xx} e_{zz} + e_{yy} e_{zz}
	\end{aligned}
\end{equation}

\subsection{Strain-Mode Coupling Terms, Normal Strain}

Terms relevant to $Q_1$

\begin{equation}
	\begin{aligned}
		E_1(e Q_1^2) &= e_{xx} Q_{1x}^2 + e_{yy} Q_{1y}^2 + e_{zz} Q_{1z}^2 \\
		E_2(e Q_1^2) &= e_{xx} Q_{1y}^2 + e_{xx} Q_{1z}^2 + e_{yy} Q_{1x}^2 + e_{yy} Q_{1z}^2 + e_{zz} Q_{1x}^2 + e_{zz} Q_{1y}^2
	\end{aligned}
\end{equation}

Terms relevant to $Q_4$

\begin{equation}
	\begin{aligned}
		E_1(e Q_4^2) &= e_{xx} Q_{4x}^2 + e_{yy} Q_{4y}^2 + e_{zz} Q_{4z}^2 \\
		E_2(e Q_4^2) &= e_{xx} Q_{4y}^2 + e_{xx} Q_{4z}^2 + e_{yy} Q_{4x}^2 + e_{yy} Q_{4z}^2 + e_{zz} Q_{4x}^2 + e_{zz} Q_{4y}^2
	\end{aligned}
\end{equation}

Terms relevant to $Q_6$

\begin{equation}
	\begin{aligned}
		E_1(e Q_6^2) &= e_{xx} Q_{6yx}^2 + e_{xx} Q_{6zx}^2 + e_{yy} Q_{6xy}^2 + e_{yy} Q_{6zy}^2 + e_{zz} Q_{6xz}^2 + e_{zz} Q_{6yz}^2 \\
		E_2(e Q_6^2) &= e_{xx} Q_{6xy}^2 + e_{xx} Q_{6xz}^2 + e_{yy} Q_{6yx}^2 + e_{yy} Q_{6yz}^2 + e_{zz} Q_{6zx}^2 + e_{zz} Q_{6zy}^2 \\
		E_3(e Q_6^2) &= e_{xx} Q_{6yz}^2 + e_{xx} Q_{6zy}^2 + e_{yy} Q_{6xz}^2 + e_{yy} Q_{6zx}^2 + e_{zz} Q_{6xy}^2 + e_{zz} Q_{6yx}^2
	\end{aligned}
\end{equation}

\subsection{Strain-Mode Coupling Terms, Shear Strain}

\begin{equation}
	\begin{aligned}
		E_3(e Q_4^2) &= e_{xy} Q_{4x} Q_{4y} + e_{yz} Q_{4y} Q_{4z} + e_{zx} Q_{4x} Q_{4z} \\
		E_4(e Q_6^2) &= e_{xy} Q_{6zx} Q_{6zy} + e_{yz} Q_{6xy} Q_{6xz} + e_{zx} Q_{6yx} Q_{6yz}
	\end{aligned}
\end{equation}

\section{Possible Ferroelectric Topological Domains}

Though the stable condition of DWs is summarized from the DW model with two domains, we can still explore the existence of ferroelectric topological domains. 
If the DW interfaces in the ferroelectric topological domain are stable, then the ferroelectric topological domain is likely to be stable. 
It is worth to mention that this condition that the interfaces are all stable using the stable condition summarized from DW model is a necessary condition but not a sufficient condition, therefore, there are more stable ferroelectric topological domains than that we predicted. In addition, defects including dopants and vacancies can stabilize topological domains that are not stable without defects. 

The following figures \ref{fig:topological-domain-1} and \ref{fig:topological-domain-2} show some of the ferroelectric vortex structures which predicted to be stable. These structures exhibit the possible combinations of pseudo-chirality numbers that can stabilize all the interfaces. We did not do structure relaxations on them, because they don't satisfy the periodic boundary condition and are typically very large, which makes it hard for first-principle calculation. 

\begin{figure}
    \includegraphics[scale=0.7]{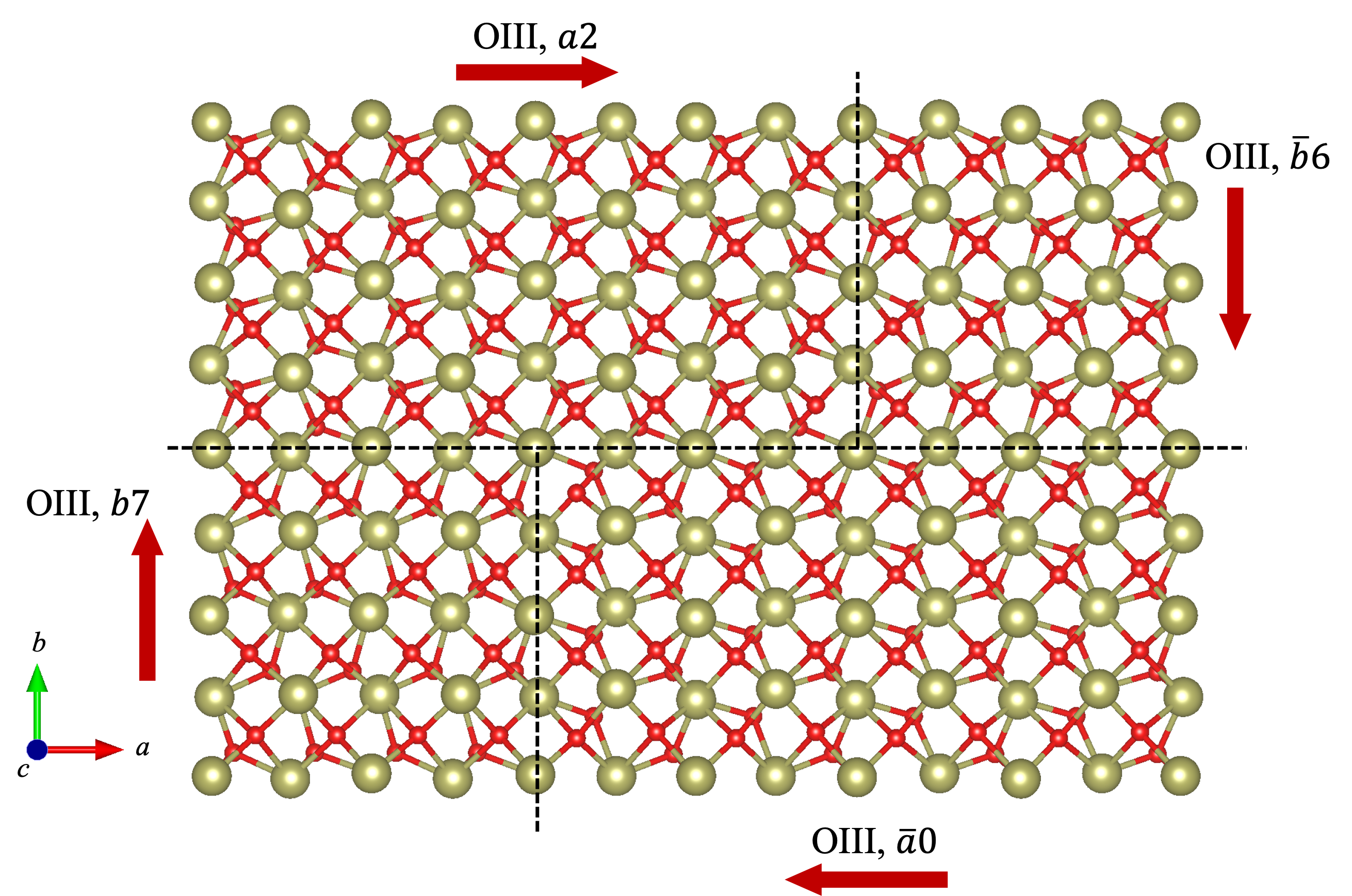}
    \caption{Ferroelectric Vortex (I). 
	Red arrows indicate polarization directions of domains. 
    Black wavy lines indicate domain interfaces. 
    \label{fig:topological-domain-1}}
\end{figure}

\begin{figure}
    \includegraphics[scale=0.7]{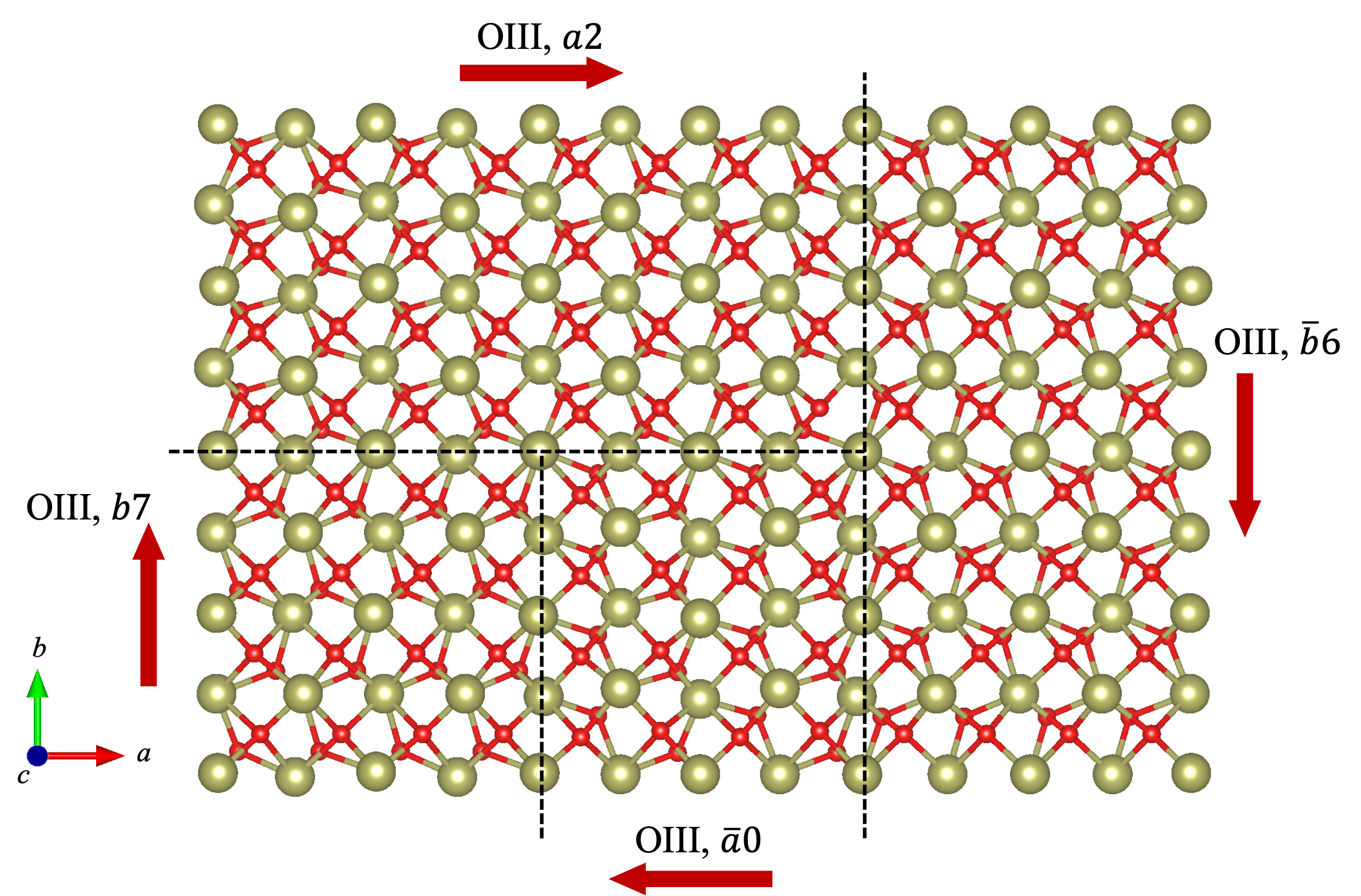}
    \caption{Ferroelectric Vortex (II). 
	Red arrows indicate polarization directions of domains. 
    Black wavy lines indicate domain interfaces. 
    \label{fig:topological-domain-2}}
\end{figure}

\end{document}